\documentclass[pra,superscriptaddress,twocolumn,floatfig,showpacs,floatfix]{revtex4-1}
\usepackage{graphicx,color}
\usepackage{amsmath,amssymb,bm}
\usepackage{braket}		% Dirac notation

\definecolor{darkred}{rgb}{0.90,0,0}
\definecolor{darkgreen}{rgb}{0,0.60,.2}
\definecolor{darkblue}{rgb}{0,0,1}
\definecolor{grey}{cmyk}{0,0,0,0.25}
\definecolor{orange}{cmyk}{0,0.6,1,0}

\begin{document}
\title{Real-time decay of a highly excited charge carrier in the one-dimensional Holstein model
}

\author{F. Dorfner}
\author{L. Vidmar}
\email[E-mail: ]{Lev.Vidmar@lmu.de}
\affiliation{Department of Physics and Arnold Sommerfeld Center for Theoretical Physics,
Ludwig-Maximilians-Universit\"at M\"unchen, D-80333 M\"unchen, Germany}
\author{C. Brockt}
\author{E. Jeckelmann}
\affiliation{Institut f\"{u}r Theoretische Physik, Leibniz Universit\"{a}t Hannover, Appelstrasse 2, D-30167 Hannover, Germany}
\author{F. Heidrich-Meisner}
\affiliation{Department of Physics and Arnold Sommerfeld Center for Theoretical Physics,
Ludwig-Maximilians-Universit\"at M\"unchen, D-80333 M\"unchen, Germany}

\begin{abstract}
We study the real-time dynamics of a highly excited charge carrier coupled to quantum phonons via a Holstein-type electron-phonon coupling.
This is a prototypical example for the non-equilibrium dynamics in an interacting many-body system where excess energy is transferred from electronic to phononic degrees of freedom.
We use diagonalization in a limited functional space (LFS) to study the non-equilibrium dynamics on a finite one-dimensional chain.
This method agrees with exact diagonalization and the time-evolving block-decimation method, in both the relaxation regime and the
long-time stationary state, and among these three methods it is the most efficient and versatile one for this problem.
We perform a comprehensive analysis of the time evolution by calculating the electron, phonon and electron-phonon coupling energies, and the electronic momentum distribution function.
The numerical results are compared to analytical solutions for short times, for a small hopping amplitude and for a weak electron-phonon coupling.
In the latter case, the relaxation dynamics obtained from the Boltzmann equation agrees very well with the LFS data.
We also study the time dependence of the eigenstates of the single-site reduced density matrix, which defines so-called optimal phonon modes.
We discuss their structure in non-equilibrium and the distribution of their weights.
Our analysis shows that the structure of optimal phonon modes contains very useful information for the interpretation of the numerical data.
\end{abstract}

\date{Draft of \today}

\maketitle

% 75.10.Pq  Magnetic ordering, general theory and models of spin chain models
% 71.27.+a  Strongly correlated electron systems
% 75.40.Mg  Computer modeling and simulation of magnetic critical points
% 05.60.Gg  Transport processes, quantum
%%%%%%%%%%%%%%%%%%%%%%%%%%%%%%%%%%%%%%%%%%%%%%%%%%%%%%%%%%%%%%%%%%%%%%%%%%%%%%%

\section{Introduction}
\label{sec:intro}

The energy transfer from charge carriers to their environment represents one of
the focal points of research in the dynamics of condensed-matter systems.
The excitation of the electronic subsystem and the measurement of its relaxation in real time may provide valuable insight into fundamental processes in strongly correlated systems~\cite{orenstein12}, and it may also be exploited in energy conversion devices such as solar cells~\cite{sheng14}.
The interest in non-equilibrium systems with strong correlations has recently been intensified by a rapid development of time-resolved experiments, which enable one to follow the evolution of the excited system from the early stage (of the order of a few electronic time units~\cite{dalconte15,gadermaier10,novelli14}) up to the final, thermalized state.
One of the central mechanisms of relaxation is the energy transfer from charge carriers to phonons since the latter type of excitations is ubiquitous in solids.

The precise role of phonons in dynamics is nevertheless a very subtle issue that depends on many details of the underlying many-body system.
In some cases, the coupling of phonons to the charge carriers is so strong that polaronic effects represent the major many-body effect in the dynamics.
These effects were studied experimentally in the context of polaron formation (or more generally, self-trapping of charge carriers)~\cite{ge98,tomimoto98,dexheimer00,sugita01,miller02,gahl02,dean11,morrissey13}, and polaron transport~\cite{scherff13,ouellette13,sheu14}.
Phonons also represent a key ingredient to understand the dynamics of photo-excited charge carriers in semiconductors~\cite{rossi02}.
In addition, even if electronic correlations strongly influence the many-body spectrum, phonons may still provide the dominant (and the fastest) relaxation channel.
For example, the key role of phonons for relaxation has been conjectured for one-dimensional (1D) Mott insulators~\cite{matsueda12,mitrano14} (where the spin relaxation channel is inefficient~\cite{al-Hassanieh2008} due to spin-charge separation), and has been discussed in the context of the mechanism of ultrafast demagnetization of ferromagnets~\cite{koopmans10,carva13}.
On the contrary, recent results for two-dimensional systems with antiferromagnetic correlations have shown that relaxation due to coupling to antiferromagnetic spin excitations can be very fast~\cite{golez14,iyoda14,eckstein14a,eckstein14b}, and on a short time scale, these excitations can absorb more energy than phonons~\cite{vidmar11c}.

To clarify the role of phonons in non-equilibrium systems in more detail, the main theoretical questions that motivate our investigation are:
{\it (i)} How efficient is the energy transfer to phonons, depending on the characteristic energy scales of the electrons and phonons and the electron-phonon coupling strength?
{\it (ii)} What is the relevant time scale for the energy transfer to phonons?
{\it (iii)} What is the relation between closed and open systems with respect to dissipation and what is the dynamics beyond the weak-coupling regime?
In connection with that, to which extent is the knowledge of the unitary time evolution required to describe the dynamics of a quantum many-body system, or in which cases are semi-classical approaches sufficient?
{\it (iv)} How to numerically treat quantum many-body systems with bosonic degrees of freedom far away from equilibrium?

Various directions of the recent developments of non-equilibrium techniques offer a rich perspective to answer the last question.
However, before the era of high-precision time-resolved experiments, most of the methods were developed to address transport properties within semi-classical approaches (see, e.g., Ref.~\cite{emin87}).
Methods retaining full quantum coherence have recently been developed to study transport~\cite{bonca97,vidmar11,cheung13} and relaxation~\cite{ku2007,fehske11,golez12b,li13} of isolated carriers coupled to phonons.
(Note that exact solutions can be obtained in the 1D adiabatic regime and for the linear electronic dispersion~\cite{med96,fricke97}.)
Several techniques are also now available to study many-electron systems beyond the mean-field~\cite{krull14} and Boltzmann approaches~\cite{kabanov2008,baranov14}.
For example, dynamics of electrons coupled to phonons has been studied within the Holstein model using dynamical mean-field theory (DMFT)~\cite{mura14}, continuous-time quantum Monte Carlo~\cite{hohenadler13} and Keldysh Green functions within the Migdal approximation~\cite{kemper13,sentef13}.
In addition, by extending the problem to electronic correlations, studies of the Hubbard-Holstein and the $t$-$J$ Holstein model have been performed using methods based on exact diagonalization~\cite{vidmar11c,yonemitsu09,defilippis12,kogoj14}, the density-matrix renormalization group~\cite{matsueda12} and dynamical mean-field theory~\cite{werner13,werner14}.

In this work we apply wavefunction-based methods to study the non-equilibrium dynamics of a coupled electron-phonon system.
Already on the level of ground-state calculations, the maximal number of local
phonons $N_{\rm max}$ has been identified as the bottleneck for efficient simulations.
An efficient truncation of the phonon basis therefore represents a crucial step to overcome the rapid growth of the Hilbert space~\cite{zhang98,zhang99,weisse00}.
Density-matrix renormalization group (DMRG) algorithms can treat much larger system sizes than exact diagonalization, but they also scale unfavorably in $N_{\rm max}$ (see, e.g., Refs.~\cite{jeckelmann99,ejima09} for the Holstein model and Refs.~\cite{fehske04,tezuka05,fehske08,nocera14} for the Hubbard-Holstein model).
Here we focus on a single charge carrier, which, as the main advantage, allows for the numerically reliable treatment of the time evolution in a broad parameter regime.
We study the Holstein Hamiltonian
\begin{eqnarray}
H & = & - t_0 \sum_j \left( c_j^{\dag} c_{j+1}^{\phantom{\dag}} + c_{j+1}^{\dag}
c_j^{\phantom{\dag}} \right) + \hbar \omega_0 \sum_j b_j^{\dag}
b_j^{\phantom{\dag}}  \\
&& - \gamma \sum_j \left( b_j^{\dag} + b_j^{\phantom{\dag}} \right) n_j, \nonumber
\end{eqnarray}
where $b_j$ ($c_j$) annihilates a phonon (electron) at site $j$ and the local electronic density on site $j$ is
$n_j = c_j^{\dag} c_j^{\phantom{\dag}}$.
We set $\hbar \equiv 1$ throughout the paper.

\begin{figure}[!t]
\includegraphics[width=.60\columnwidth]{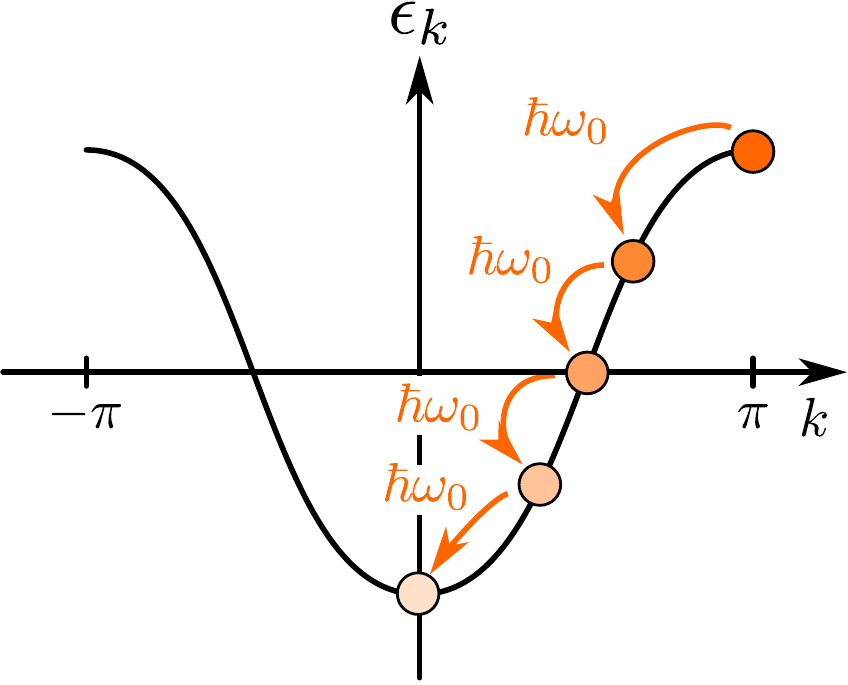}
\caption{(Color online)
Sketch of the initial condition, Eq.~(\ref{psi0}), and the time evolution:
We start from the state with one electron at momentum $k=\pi$ and no phonon.
The total energy is thus equal to the initial electronic kinetic energy $E_{\rm{kin}}(t=0) = \epsilon_\pi = 2 t_0$, where $\epsilon_k = -2t_0 \cos{k}$.
Due to the coupling to phonons, the electron loses energy by exciting phonons while moving through the lattice, which also results in the redistribution of its momentum occupations.
}
\label{fig:sketch}
\end{figure}

We pursue three main goals.
First, we compare different wavefunction-based methods, i.e., exact diagonalization, diagonalization in a limited functional space~\cite{bonca99} and the time-evolving block-decimation (TEBD) method~\cite{vidal2004}, which all show perfect agreement.
We find that the most powerful method to treat this class of problem is diagonalization in a limited functional space, first introduced by Bon\v{c}a {\it et al.}~\cite{bonca99} to describe the Holstein polaron ground state.
We apply the method on a finite lattice and show that it allows for the efficient simulation of dynamics in {\it both} the relaxation regime as well as in the long-time stationary regime.
This complements a previous work using the same method~\cite{golez12b}, where the relaxation regime on an infinite lattice was studied.

Second, we want to analytically and numerically describe the non-equilibrium dynamics in limiting cases and in the crossover from weak to strong electron-phonon coupling, as well as in the crossover from adiabatic to anti-adiabatic regime.
To be specific, we are interested in the dynamics emerging from an initial state
with all excess energy contained in the electronic sector
\begin{equation} \label{psi0}
\ket{\psi_0} = c_K^\dagger \ket{\emptyset}_{\rm ele} \otimes \ket{\emptyset}_{\rm ph}, 
\end{equation}
where $c_k^\dagger$ represents the creation operator for an electronic state with momentum $k$ (see also Fig.~\ref{fig:sketch}).
We choose $k=\pi$, which also sets the total crystal momentum $K$ of the coupled electron-phonon system.
The two states in Eq.~\eqref{psi0} represent the electron and the phonon vacuum, respectively.
This state is an eigenstate of the system when the electron-phonon coupling is zero and is one of the simplest initial states where the charge carrier is highly excited.
The physical motivation for choosing the state in Eq.~(\ref{psi0}) is to model
the dynamics of a free-electron wave-packet~\cite{ku2007,fehske11,li13} emerging after a sudden external perturbation of a many-body system. 
Preventing density fluctuations (i.e., by using the fully delocalized initial state with a sharp momentum) simplifies the analysis in the stationary regime, discussed in Sec.~\ref{subsec:steady}, since our state is homogeneous for all times.
We define two distinct regimes of the time evolution:
{\it (i)} The {\it relaxation regime}, which is characterized by a net transfer of energy between the electronic and phononic system.
For some parameter regimes, coherent oscillations with the period $2\pi/\omega_0$ emerge in the dynamics.
In such cases, the relaxation regime is defined by a nonzero net energy transfer averaged over one oscillation period.
However, to have a meaningful definition of relaxation and energy transfer, we need to require that the amplitude of oscillations is sufficiently small compared to the mean value of energy transfer.
{\it (ii)} The {\it stationary regime}, in which no net energy transfer is observed.
This definition covers both the case where the observables indeed become time independent, as well as the case with persisting coherent oscillations (i.e., when the amplitude of oscillations does not decay to zero).
Such terminology differs from other thermalization studies where temporal fluctuations about the time average are usually required to be arbitrarily small to obtain a stationary state (see, e.g., Refs.~\cite{rigol08,polkovnikov11,langen14,eisert14} and references therein).

As a third goal, we extract the single-site reduced density matrix from the time-dependent total wavefunction.
Motivated by the idea of the DMRG algorithms~\cite{white92,schollwock2005density,schollwock2011density},
we calculate the von Neumann entropy and the eigensystem of the single-site density matrix.
In the electron-phonon coupled systems, the corresponding eigenstates with the largest eigenvalues represent the {\it optimal phonon modes}.
Zhang {\it et al.}~\cite{zhang98,zhang99} first used the optimal phonon modes to
truncate the phonon Hilbert space self-consistently in a study of the Holstein model in equilibrium.
Later, this idea was applied to the spin-Peierls model~\cite{Friedman2000} as
well as to the Holstein-Hubbard model~\cite{weisse00,weisse02}.
Here we analyze the optimal phonon modes in a non-equilibrium set-up. We do not
use them as a tool for truncating the Hilbert space, but rather to gain additional insight into the dynamics.
We show that the structure of the optimal phonon modes carries valuable information about physical processes, and may complement the analysis based on observables only.

{\it Summary of the results.}
One of the central measures of whether the energy transfer between the electron and phonons is substantial or not is the adiabaticity parameter, i.e., the ratio between the phonon energy $\omega_0$ and the electronic bandwidth, set by the electronic hopping amplitude $t_0$.
If this ratio is sufficiently small, relaxation dynamics can be observed.
We construct a set of Boltzmann equations to describe the relaxation and compare the results to the numerical data at weak electron-phonon coupling, obtaining a very good agreement.
The Boltzmann equation also provides a convenient framework to analytically derive the characteristic relaxation time $\tau$ for a constant density of states.
Its quantitative value is given by $\tau \omega_0 = (16/\pi) (\gamma/t_0)^{-2}$, and interestingly, it also reasonably well describes the numerical results obtained for the
density of states of a one-dimensional tight-binding system.  

On the contrary, in the anti-adiabatic limit $\omega_0 \gg t_0$
the dynamics in both weak- and strong-coupling regimes is governed by coherent oscillations with the period $2\pi/\omega_0$ and a negligible energy transfer.
We have applied perturbation theory in both limits $\gamma \ll t_0$ and $\gamma \gg t_0$ and obtained very similar results for the time dependence of observables.
The paradigmatic case to describe the oscillatory dynamics is the single-site problem $t_0 = 0$: here, the time evolution of observables can be obtained analytically for all times.
While the frequency of oscillations is given by $\omega_0$, the amplitude of the oscillations is set by the ratio $\gamma/\omega_0$.

We then numerically calculate the time-dependent expectation values of observables in the crossover from adiabatic to anti-adiabatic regime as well as from weak to strong electron-phonon coupling.
The generic evolution of the electronic momentum-distribution function is consistent with the sketch presented in Fig.~\ref{fig:sketch}.
It undergoes strong redistributions for the initial state in Eq.~(\ref{psi0}):
initially, all the weight is located at $k=\pi$, while at later times a maximum
at $k=0$ develops, accompanied by a reduction of the electronic kinetic energy.
The stationary regime is typically characterized by the maximum of the momentum-distribution function being at $k=0$ and persistent coherent oscillations of the observables about their average; however, we show that the oscillations vanish with increasing lattice size for sufficiently weak electron-phonon coupling and away from the anti-adiabatic limit.
Note that both the electronic kinetic energy and the electron-phonon coupling energy are lower in the stationary regime in comparison to its initial value, hence both contributions should be accounted for when quantifying the energy redistribution.
We construct a heuristic estimate, supported by the numerical data, of the average number of emitted phonons in the stationary regime:
they equal the total energy minus the sum of the kinetic and the coupling energy in the corresponding ground state.
The time evolution can therefore simply be viewed as the energy transfer from the subsystem, containing both the electronic and the electron-phonon coupling energy, to the phononic subsystem.

We finally calculate the entanglement entropy and the eigensystem of the single-site reduced density matrix to investigate how the optimal phonon modes evolve in time.
For the optimal phonon modes with the largest weights, the weight distribution
remains approximately exponential in non-equilibrium as found for ground states, 
however, with a much slower decay. 
The structure of the most relevant optimal phonon modes is very interesting at strong electron-phonon coupling.
In particular, the optimal phonon mode with the largest weight exhibits time-dependent oscillations between the 
phonon-vacuum and a state with a Poisson-like distribution over phonon occupation numbers.
This structure strongly resembles the results obtained in the $t_0=0$ limit,
suggesting that the single-site coherent oscillations govern the dynamics also
when none of the model parameters is vanishingly small.

Note that the same problem, albeit with slightly different initial states and in the limit of an infinite coordination 
number, has very recently been studied by Sayyad and Eckstein using a non-equilibrium DMFT method~\cite{sayyad14}.
Their main interest is in the adiabatic regime of phonon frequencies and in an extended analysis of photoemission spectra. 
Their results for the weak-coupling regime are in agreement with our discussion.

{\it Outline.}
The plan of the paper is the following.
In Sec.~\ref{sec:setup} we introduce the model and discuss its limiting cases.
A comparison of the numerical methods, i.e., exact diagonalization, diagonalization in a limited functional space and the time-evolving block-decimation method, is performed in Sec.~\ref{sec:methods}.
We then discuss the dynamics in limiting cases by applying perturbative techniques in Sec.~\ref{sec:perturbative}, and we investigate the relaxation dynamics within the Boltzmann approach.
In Sec.~\ref{sec:results}, we study the time evolution and steady-state properties of several observables around the crossover from adiabatic to anti-adiabatic regime as well as from weak to strong electron-phonon coupling.
Finally, we focus on optimal phonon modes and the single-site entanglement entropy in Sec.~\ref{sec:optmodes}.
We present our conclusions in Sec.~\ref{sec:conclusion}.

%%%%%%%%%%%%%%%%%%%%%%%%%%%%%%%%%%%%%%%%%%%%%%%%%%%%%%%%%%%%%%%%%%%%%%%%%%%%%%%
\section{The Holstein model}
\label{sec:setup}

The Holstein model~\cite{holstein1959325} is a widely studied model in
condensed matter theory, mimicking the interaction of charge carriers with local lattice vibrations.
Recently, the possibility of the quantum simulation of the Holstein model has been discussed in the context of ultracold bosons~\cite{bruderer07}, polar molecules~\cite{herrera11,herrera13}, trapped ions~\cite{stojanovic12,mezzacapo12}, Rydberg atoms~\cite{hague12} and superconducting circuits~\cite{mei13}.
The Holstein model consists of three terms
\begin{equation}
H = H_{\rm{kin}} + H_{\rm{ph}} +  H_{\rm{coup}}, \label{holstein_ham}
\end{equation}
which represent the electronic kinetic energy $H_{\rm{kin}}$, the phonon energy $H_{\rm{ph}}$ and the electron-phonon coupling energy $H_{\rm{coup}}$.
We provide these terms both in real- and momentum-space representations
\begin{eqnarray}
H_{\rm{kin}} &=& - t_0 \sum_j \left( c_j^{\dag} c_{j+1}^{\phantom{\dag}} +
c_{j+1}^{\dag} c_j^{\phantom{\dag}} \right) = \sum_k \epsilon_k c_k^{\dag}
c_k^{\phantom{\dag}}, \label{hkin} \\
H_{\rm{ph}} &= & \omega_0 \sum_j b_j^{\dag} b_j^{\phantom{\dag}} =  \omega_0
\sum_q b_q^{\dag} b_q^{\phantom{\dag}}, \label{hph} \\
H_{\rm{coup}} &=& - \gamma \sum_j \left( b_j^{\dag} + b_j^{\phantom{\dag}} \right) n_j \label{hcoup} \\ \nonumber
& = &  - \frac{\gamma}{\sqrt{L}} \sum_{q,k} \left(  b_q^{\dag} c_{k-q}^{\dag}
c_k^{\phantom{\dag}} + \text{h.c.}  \right).
\end{eqnarray}
We are interested in a single-electron system only, hence the filling is kept at $n = 1/L$ where $L$ is the number of sites on a 
1D lattice.
Unless stated otherwise, periodic boundary conditions are assumed.
In the momentum-space representation, the annihilation operator of a particle with momentum $k$ is defined by a discrete Fourier transformation
$d_k = \sum_j e^{-ijk} d_j / \sqrt{L}$, where $d_j \in \{ b_j,c_j \}$.
The electron in Eq.~(\ref{hkin}) has the standard tight-binding dispersion
\begin{equation} \label{e_dispersion}
\epsilon_k = -2t_0 \cos{k},
\end{equation}
while the phonons in Eq.~(\ref{hph}) are dispersionless.
We also study the electronic momentum distribution function $n_k$, which is defined as
\begin{equation}
n_k = \langle c_k^\dag c_k^{\phantom{\dag}} \rangle = \frac{1}{L} \sum_{j,l} e^{i(j-l)k}
\langle c_j^{\dag} c_l^{\phantom{\dag}}\rangle .
\end{equation}
Throughout the manuscript we will denote the expectation values of $H$, $H_{\rm ph}$, $H_{\rm{coup}}$ and $H_{\rm{kin}}$ as $E_{\rm total}$, $E_{\rm ph}$, $E_{\rm{coup}}$ and  $E_{\rm{kin}}$, respectively.

In the ground state of the model at non-zero electron-phonon coupling $\gamma$, the electron and a cloud of phonons are spatially correlated and they form a composite particle with a renormalized effective mass.
This quasi-particle is called a polaron~\cite{landau1933,devreese09}.
There is no analytical solution for the Holstein polaron in the full parameter regime, however, results from numerical simulations~\cite{fehske2007} and perturbative limits may be very instructive~\cite{barisic08}.
The limit of weak electron-phonon coupling $\gamma \ll t_0$ leads to a ground state with a highly mobile electron and very few phonons.
Because of the large spatial extent of electron-phonon correlations, this case is called the \emph{large-polaron} limit.
On the contrary, the strong-coupling anti-adiabatic limit $\gamma, \omega_0 \gg
t_0$ is characterized by a reduced mobility and a rapid decay of electron-phonon
correlations with distance.
Hence it as also referred to as the \emph{small-polaron} limit.
In the extreme limit of zero hopping amplitude $t_0=0$ one is left with a fully local Hamiltonian.
This limit is particularly instructive since the Hamiltonian
\begin{align}
H = \omega_0 \sum_j b_j^{\dag} b_j^{\phantom{\dag}}  - \gamma \sum_j \left(
b_j^{\dag} + b_j^{\phantom{\dag}}  \right) n_j \label{sec2:effham}
\end{align}
can be diagonalized analytically by shifting the phonon operators at the site $j$ of the electron via
\begin{equation} \label{btransf}
b_j = a_j + g,
\end{equation}
which leads to the diagonal Hamiltonian
\begin{equation} \label{ht0}
H_j = \omega_0 \; a_j^\dag a_j^{\phantom{\dag}}  - \varepsilon_b.
\end{equation}
Here, $\varepsilon_b$ represents the polaron binding energy
\begin{equation}
\label{polaron_energy}
\varepsilon_b = \frac{\gamma^2}{\omega_0}
\end{equation}
and we have introduced the shift given by
\begin{equation} \label{gdef}
g = \frac{\gamma}{\omega_0},
\end{equation}
which emerges as the only relevant model parameter in the $t_0 \to 0$ limit.
The ground-state wavefunction in the translationally invariant case can be written as 
\begin{equation}
\label{eq:gs}
\ket{\psi_{\rm GS}} = \frac{e^{-g^2/2}}{\sqrt{L}} \sum_j 
\left [e^{g b_j^\dagger} \ket{\emptyset}_{\rm ph}  
\otimes c_j^{\dag} \ket{\emptyset}_{\rm ele} \right ] .
\end{equation}
The average number of phonons in this state is $N_{\rm ph} =\sum_j \langle b_j^\dagger b_j \rangle = g^2$.
The transformation of operators defined by Eq.~(\ref{btransf}) corresponds to a unitary transformation to the new orthogonal basis, known also as the coherent-state basis.
We will elaborate in more detail on this transformation in Sec.~\ref{sssec:sclimit} where we derive an analytical expression for the real-time dynamics in the limit of small $t_0$.
In addition, the coherent-state basis also represents a very instructive framework for the discussion of optimal modes in Sec.~\ref{sec:optmodes}.

The intermediate regime of parameters is characterized by a crossover from the large to the small polaron case with increasing electron-phonon coupling.
The properties are usually characterized as a function of the dimensionless electron-phonon coupling parameter
\begin{align} \label{def_lambda}
\lambda &= \frac{\gamma^2}{2t_0\omega_0} = \frac{\varepsilon_b}{2t_0}.
\end{align}
This is the ratio between the ground-state energies in two extreme cases:
the free fermion energy $-2t_0$, which represents the ground-state energy at zero
electron-phonon coupling, and the polaron binding energy $-\varepsilon_b$, which represents the ground-state energy at infinitely strong electron-phonon coupling.
Hence the crossover (i.e., the point where most observables exhibit the most rapid variation) emerges roughly at $\lambda^* \approx 1$.
The precise crossover value also depends on another dimensionless parameter 
$\alpha = \omega_0/t_0$~\cite{alvermann10}, called the adiabaticity parameter, which is used to distinguish the adiabatic limit ($\alpha \ll 1$) from the anti-adiabatic limit ($\alpha \gg 1$).
It gives a measure for the stiffness of the lattice: in the adiabatic limit the lattice belatedly adjusts to the electronic motion, while in the anti-adiabatic limit it adjusts almost instantaneously.
In the context of relaxation dynamics, the adiabaticity parameter determines whether the energy transfer is substantial or not.
The difference will be illustrated in Sec.~\ref{sec:perturbative}.
While the energy transfer is substantial in the weak-coupling adiabatic regime (c.f. Sec.~\ref{subsec:boltzmann}), it is inefficient or even totally absent in the anti-adiabatic regime (c.f. Secs.~\ref{sssec:wanti} and~\ref{sssec:sclimit}).

%%%%%%%%%%%%%%%%%%%%%%%%%%%%%%%%%%%%%%%%%%%%%%%%%%%%%%%%%%%%%%%%%%%%%%%%%%%%%%%
\section{Numerical methods} \label{sec:methods}

Electron-phonon lattice models such as the Holstein model~\eqref{holstein_ham} do not conserve the number of phonons in the system and the Hilbert space is infinite even for a system on a finite lattice.
Wavefunction based methods used to study the Holstein polaron (e.g., exact diagonalization~\cite{marsiglio95,wellein96,wellein97,weisse06} and DMRG~\cite{jeckelmann98}) are flexible and can yield 
essentially exact results, however, they face an obvious limitation since they require a finite Hilbert-space basis.
The central task is hence to efficiently construct a truncated basis that is able to represent the states one is interested in.
Typically, one truncates the total number of phonons, however, the efficiency of the truncation strongly depends on the regime of model parameters considered.
In Sec.~\ref{sec:3a} we introduce the concept of diagonalization in a limited functional space, while in Sec.~\ref{subsec:tebd} we describe the TEBD method.
Other methods widely used to study properties of the Holstein polaron are dynamical mean-field theory~\cite{ciuchi97}, different versions of quantum Monte Carlo (e.g., diagrammatic~\cite{prokofev98,goodvin11}, continuous-time~\cite{kornilovitch98,hague06}, Lang-Firsov based~\cite{hohenadler04} and variational Monte Carlo~\cite{ohgoe14}), and a momentum-average approach~\cite{berciu06,goodvin06,barisic07,berciu07,goodvin11}.

\subsection{Diagonalization in a limited functional space}
\label{sec:3a}

The main idea in this truncation scheme comes from the structure of the polaron: Since the electron and phonons are correlated in space, the phonons in the vicinity of the electron are more relevant than the phonons far away from the electron.
One should therefore find a way to keep a large amount of distinct phonon configurations around the electron while neglecting some states with less important phonon configurations.
This is efficiently achieved by constructing a limited functional space (LFS), first introduced by Bon\v{c}a {\it et al.}~\cite{bonca99}.
Instead of the full basis, it only picks up a limited set of wavefunctions, i.e., configurations in the occupation-number basis, represented in a translationally invariant form.
For a given size of the functional space, the system is then diagonalized exactly.
The method has been shown to be both very accurate and efficient, and it was applied to many different problems~\cite{bonca00,bonca07,bonca08,vidmar09,vidmar10,vidmar11,vidmar11c,golez12a,golez12b} (see also~\cite{barisic02,barisic04,alvermann10,li10,li13,chakraborty13}).

\begin{figure}[!t]
\includegraphics[width=.96\columnwidth]{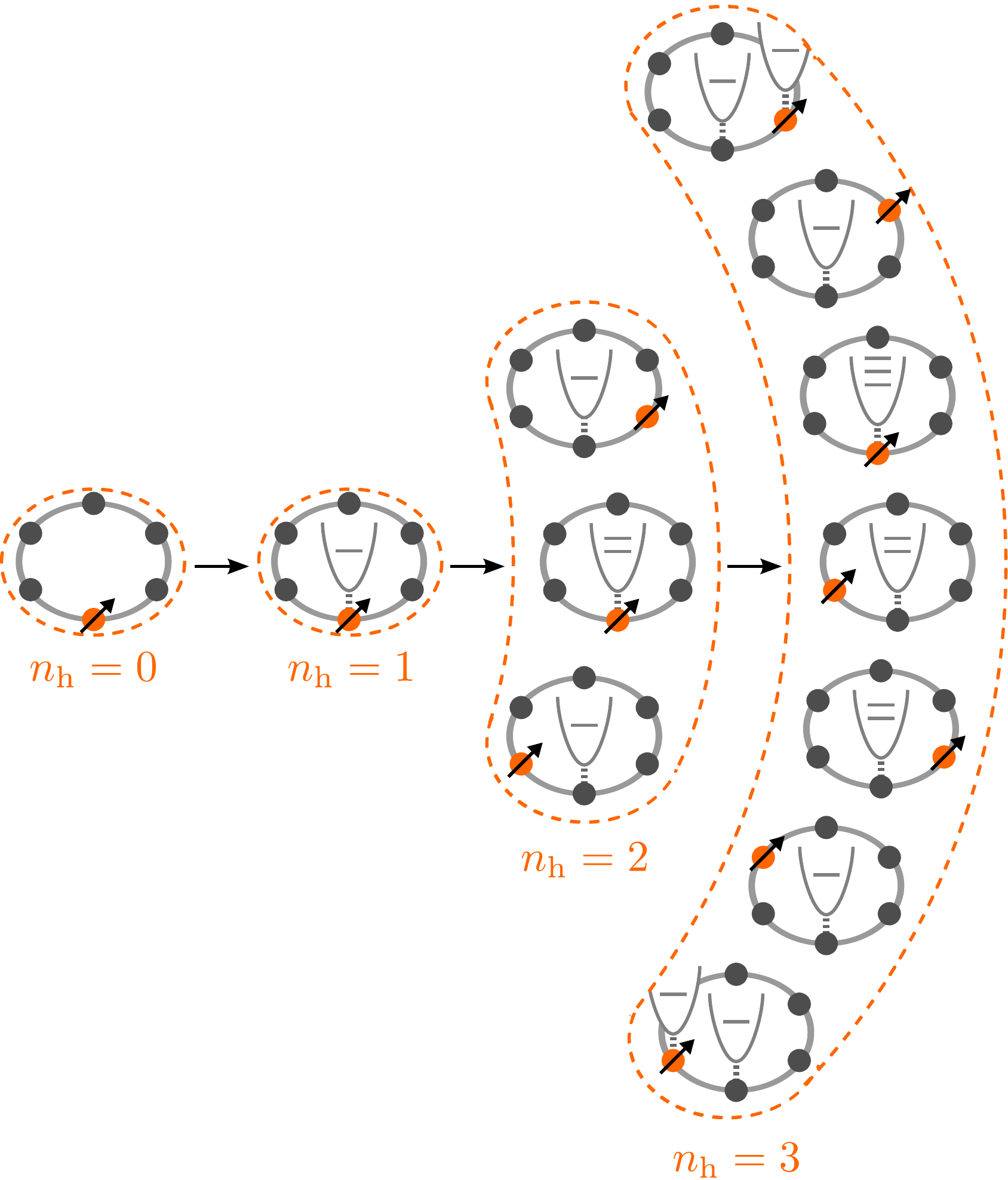}
\caption{(Color online)
Sketch of the states in the limited functional space that are generated by Eq.~(\ref{def_lfs}) during the first three iterations (we take $M=1$ for simplicity).
The ring represents a lattice with $L=6$ sites and periodic boundary conditions.
Each state in the figure represents the parent state of a translationally invariant state.
The site of the electron is represented by an arrow for clarity, while the local phonons are represented by harmonic oscillators.
Starting from the free electron state ($n_{\rm h} = 0$), a single phonon (since $M=1$) is created on the site of the electron.
In the next generation, the $n_{\rm h}=1$ state is taken as the starting state: the electron hops to the left or right of the phonon (leading to two new states) and a state with an additional phonon is created.
Next, the second generation ($n_{\rm h} = 2$) states serve as starting states
and the new unique states of the third generation ($n_{\rm h}=3$) are added to the basis.
This procedure is repeated until the desired number of generations are set up (when a state is generated that already exists in an earlier generation, it is omitted from the generation).
%The upper bound for the number of states (when every generated state would be unique) scales as $\mathcal{O}\left[ (M+2)^{N_{\rm h}-1}\right]$.
}
\label{fig:sketch_method}
\end{figure}

One of the main advantages of the method is that it provides a systematic way of generating a limited Hilbert space.
The generation procedure is initiated by a simple starting state, which is generally the bare electron in a given momentum eigenstate (i.e., in a translationally invariant state).
Then, the off-diagonal elements of the Hamiltonian are applied to the initial state.
The maximal number of generations of new states is labeled by the parameter $N_{\rm h}$.
We represent the entire set of states $ \left \{ \left \vert \phi_{k}^{(N_{\rm h},M)} \right \rangle \right  \}$ forming the LFS by a sum
\begin{align} \label{def_lfs}
\sum_{n_{\rm h}=0}^{N_{\rm h}}\left( H_{\rm{kin}} + \sum_{m=1}^M\left(H_{\rm{coup}}\right)^m \right)^{n_{\rm h}} c_k^{\dag} \ket{\emptyset}.
\end{align}
Beside $N_{\rm h}$, there is another parameter $M$ that determines the size of the LFS: it counts the number of applications of the electron-phonon coupling term within a single generation, i.e., it may increase the amount of phonons that are created in each generation.
In this work we define states on a finite lattice, which is in contrast to the initial application of the method~\cite{bonca99} where an infinite lattice was used.
The spatial extent of the phonon cloud around the electron is governed by $N_{\rm h}$.
For the states forming the LFS on a large lattice and at small $N_{\rm h}$, the largest relative distance between the electron and a phonon is $N_{\rm h}-1$ sites.
However, in our set-up on a finite lattice we typically use $N_{\rm h} > L$, hence the states with all the possible relative distances between the electron and a single phonon are incorporated in the LFS.
The generation scheme is illustrated in Fig.~\ref{fig:sketch_method} for $N_{\rm h}=3$ and $M=1$.

The method provides numerically exact results for the ground-state properties of the Holstein polaron~\cite{bonca99,ku2002}, and can be applied to different parameter regimes.
In the weak-coupling regime, the spatial extent of the ground-state electron-phonon correlations may be large.
In addition, in most excited states there are unbound phonons that  can be arbitrarily far away from the electron.
Nevertheless, by systematically increasing $N_{\rm h}$ it has been shown that an accurate description of low-lying excited states is possible~\cite{vidmar10}.
In the strong-coupling regime, the polaron is very small and has a huge amount of phonons in its vicinity.
This property can be captured by tuning the parameter $M$~\cite{bonca08} since the maximal number of phonons on the site of the electron is $N_{\rm h} \times M$.
Note that the generator of states in Eq.~(\ref{def_lfs}) does not represent the only possibility to efficiently truncate the phonon number; for alternative methods of creating distinct phonon configurations see, e.g., Refs.~\cite{cataudella04,defilippis12b}.

\begin{figure}[!t]
\includegraphics[width=.96\columnwidth]{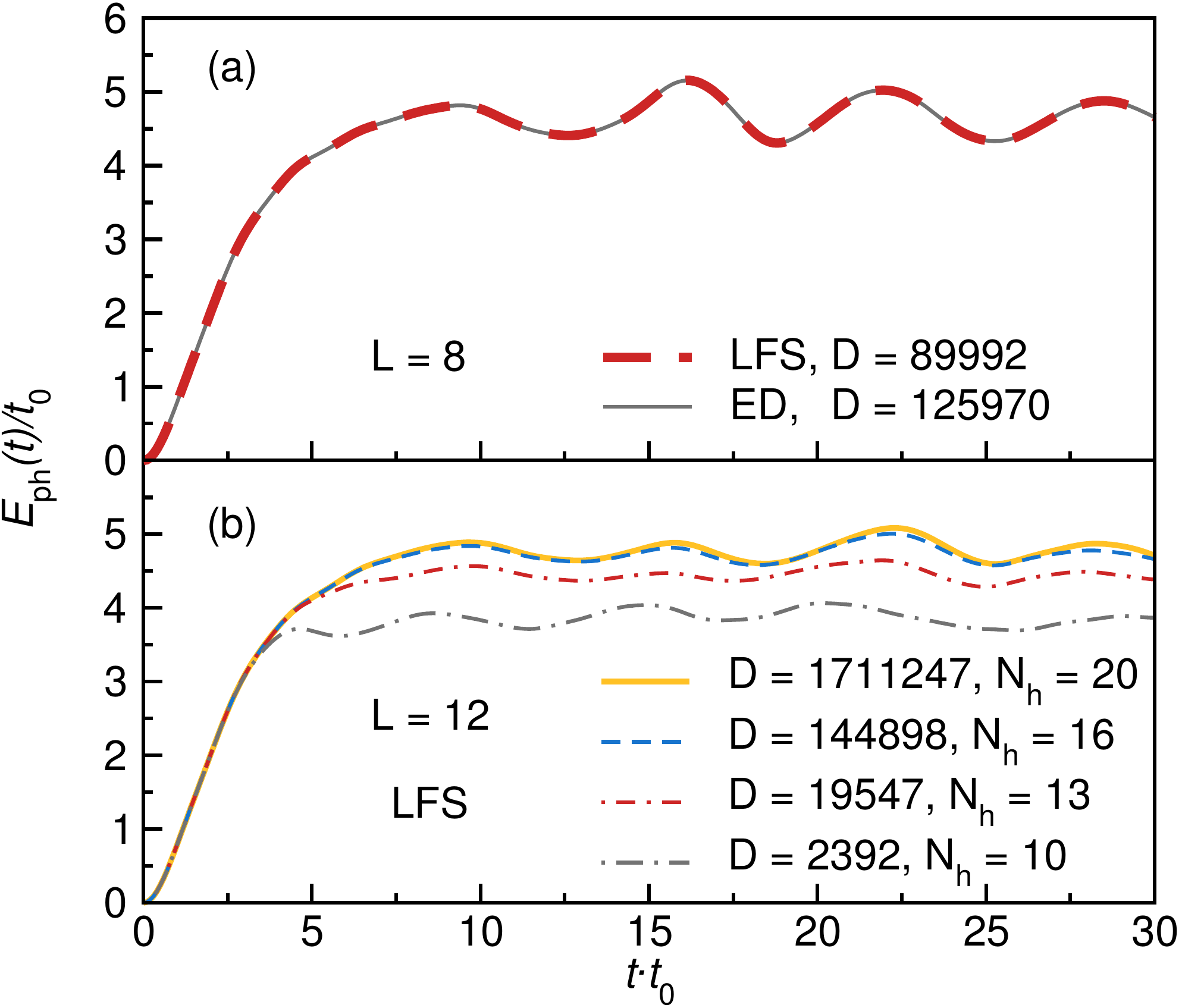}
\caption{(Color online)
Time evolution of the phonon energy $E_{\rm{ph}}(t)$ for $\lambda=0.5$ and $\omega_0 = t_0$.
(a) Comparison of the numerical methods for $L=8$.
The solid line represents results obtained by exact diagonalization (ED) in the complete Hilbert space, using the global phonon cutoff $N_{\rm{max}}=12$.
The dashed line represents results obtained by diagonalization in a limited functional space (LFS) by using the parameters $N_{\rm h}=17$ and $M=1$ [see Eq.~(\ref{def_lfs})].
(b) Convergence of $E_{\rm{ph}}(t)$ by using LFS for $L=12$.
For the parameters of the basis generation, we keep $M=1$ while $N_{\rm h}$ is varied.
The legends display the corresponding basis dimension $D$.
}
\label{fig:convergence}
\end{figure}

Recently, the LFS has also been applied to study time evolution under non-equilibrium conditions.
Two specific cases were studied: the Holstein polaron~\cite{vidmar11} or bipolaron~\cite{golez12a} driven by a constant electric field, and the Holstein polaron excited by a short pump-pulse~\cite{golez12b}.
The LFS was constructed in both cases on an infinite lattice and hence in the limit of a large time, the quasi-stationary state may include states with a diverging relative distance between the electron and the phonons.
This effect limits the largest times available by using the LFS.
Here we generalize the method to deal with a finite $L$, and use the $L \to
\infty$ limit only in a specific case in Sec.~\ref{subsec:steady}.
The benefit of this generalization is many-fold: one can, e.g., time evolve the system for a longer time interval, efficiently calculate the electronic momentum distribution function $n_k$, and analyze the $L$-dependent scaling of the amplitude of coherent oscillations in the stationary regime.

We demonstrate the efficiency of the LFS method in Fig.~\ref{fig:convergence} for the weak-coupling regime at $\lambda = 0.5$ and $\omega_0 = t_0$ by comparing to exact diagonalization in a full Hilbert space, subjected by a global phonon cutoff $N_{\rm max}$.
For both methods, we time-evolve the wavefunction $\ket{\psi(t)} = e^{-iHt} \ket{\psi_0}$ by using the iterative Lanczos algorithm~\cite{park86}.
We use a time step of $\Delta t = 0.1$ throughout the paper and the Lanczos basis is kept large enough to eliminate the numerical errors.
In Fig.~\ref{fig:convergence}(a) the result of the LFS scheme is compared to exact diagonalization using the global cutoff $N_{\rm{max}}=12$ for a system size $L=8$.
Both data are in excellent agreement.
However, when the electron-phonon coupling is further increased, a larger phonon cutoff is required, and consequently the full Hilbert space exceeds the available computational resources while the structure of the LFS can be tuned by increasing the parameter $M$ in Eq.~(\ref{def_lfs}).
Figure \ref{fig:convergence}(b) shows the convergence of the time-dependent phonon energy $E_{\rm{ph}}(t)$ at $\lambda=0.5$ with respect to the parameters $N_{\rm h}$ and $M$ of the basis generator in Eq.~(\ref{def_lfs}).
With increasing $N_{\rm h}$, the data converge fast. 
For the system size $L=12$ shown in Fig.~\ref{fig:convergence}(b), the Hilbert space of dimension $D\sim 10^6$ is sufficient for an accurate description of dynamics in all time regimes.
In all subsequent figures where the LFS results are shown, we set the parameters $N_{\rm h}$ and $M$ large enough such that the results are converged for all times.

\subsection{Time-evolving block decimation} \label{subsec:tebd}

Another method that we use is the time-evolving block-decimation (TEBD)
algorithm~\cite{vidal2004}, closely related to the time-dependent DMRG method~\cite{daley04,white04}.
The advantage of this method  is that it can also treat the case of large distances between the electron and (unbound) phonons, 
because the computational cost increases linearly with the system size for the polaron problem. 
As it is based on the matrix-product-state (MPS) formalism, this algorithm is efficient only for representing states with
small entanglement~\cite{Vidal2003}.
In this work we always start with a slightly entangled state as the matrix dimension is $m=2$ for the initial state~(\ref{psi0}), but the entanglement increases rapidly over time.
Thus we can simulate the polaron dynamics accurately only for a limited period of time, which varies with the model parameters.
In addition, we use open boundary conditions because this yields a better performance with TEBD.

\begin{figure}[!t]
\includegraphics[width=.96\columnwidth]{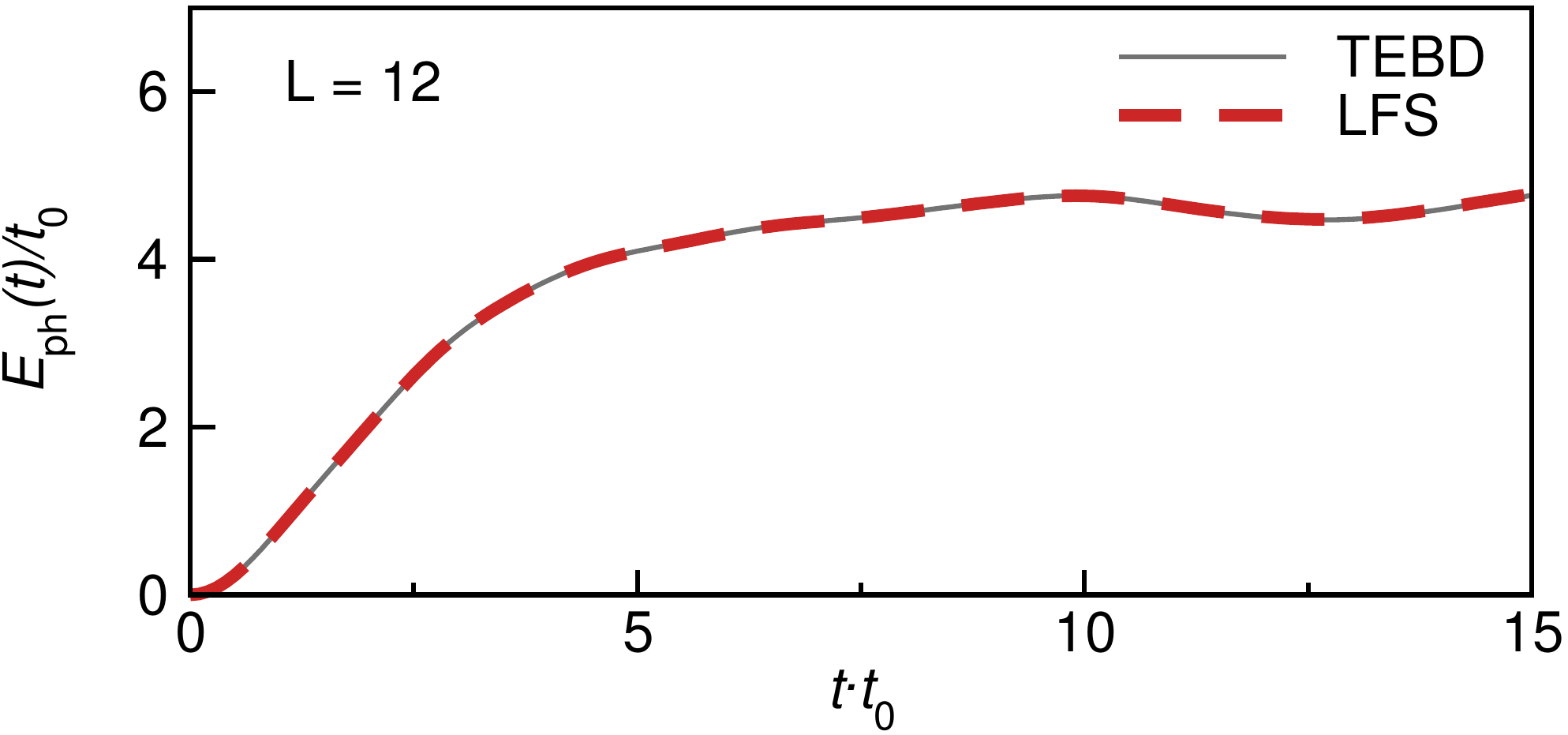}
\caption{
(Color online)
Comparison of the phonon energy $E_{\rm{ph}}(t)$ obtained by diagonalization in LFS and the TEBD method.
We set $\lambda=0.5$, $\omega_0 = t_0$ and $L=12$.
Open boundary conditions are used in this figure (we use the free-electron eigenstate with the highest energy as the initial state).
}
\label{fig:VEDTEBDComp}
\end{figure}

The TEBD method can readily be applied to systems with bosonic degrees of freedom represented by a bare boson basis with a cutoff $N_{\text{max}} \ll \infty$ as in the exact diagonalization methods~\cite{einhellinger}. 
However, since the computational cost (time) 
of TEBD scales as $N^3_{\text{max}}$, 
we have used only up to $N_{\text{max}}=17$ states per boson site and
thus we are limited to moderately large values of the coupling $\gamma \alt 2$ for the polaron problem.

In our TEBD calculations we use a second-order Trotter-Suzuki decomposition with time steps 
$\Delta t = 0.005 \omega_0$. 
In addition to the left and right normalization conditions, we also use a representation of the MPS that conserves 
the particle number explicitly.
During the time evolution 
we keep all eigenstates of the reduced density matrices that correspond to eigenvalues greater than 
$10^{-15}$ until the maximal matrix dimension ($m=50$ or $100$) is reached.
(For comparison, this corresponds to using matrix dimensions $m' = N_{\text{max}} m$ up to $m'=1700$ in TEBD simulations of 
$XYZ$ spin-$1/2$ chains.)
Thereafter the least important eigenstates are discarded but their 
weight is always smaller than $10^{-8}$.
For each time step, the resulting truncation error is smaller than the expected Trotter error.
We estimate that all TEBD results presented here have negligible errors on the figure scales.

The result of the TEBD method is compared to diagonalization within LFS in Fig.~\ref{fig:VEDTEBDComp} for $\lambda=0.5$ and 
$\omega_0 = t_0$ on a  open chain with $L=12$ sites.
The perfect agreement confirms the accuracy of both methods for long times $t$,
even for system sizes $L$ exceeding the maximal size that can be treated with exact diagonalization methods.
Since the entanglement growth makes the TEBD method computationally more expensive as time evolves and the presence of chain edges complicates the interpretation of results, 
we primarily use diagonalization in LFS in the following to comprehensively analyze the dynamics for various parameter regimes.

%%%%%%%%%%%%%%%%%%%%%%%%%%%%%%%%%%%%%%%%%%%%%%%%%%%%%%%%%%%%%%%%%%%%%%%%%%%%%%%
\section{Perturbative results} \label{sec:perturbative}

The goal of this section is to understand the dynamics in limiting cases, in which analytical expressions can be obtained.
There are two main benefits of this treatment:
{\it (i)} It provides a distinction between parameter regimes where relaxation dynamics sets in and regimes where the response is governed by coherent oscillations;
{\it (ii)} It clarifies the role of model parameters on dynamics, in particular, it determines the functional dependence of the relaxation time.

We apply time-dependent perturbation theory to obtain the time evolution of observables and compare them to the numerically exact results.
In Sec.~\ref{subsec:shorttime} we apply the general procedure to extract the lowest orders of the time-evolution operator, while in Sec.~\ref{subsec:intpicture} we carry out a perturbative expansion for the case when one parameter of the Hamiltonian is much smaller than the others.
The results of Sec.~\ref{subsec:intpicture} are applied in Secs.~\ref{subsec:weakeph} and~\ref{sssec:sclimit} to the cases of a weak electron-phonon coupling $\gamma$ and a small hopping amplitude $t_0$, respectively.
In addition, for weak electron-phonon coupling we construct the set of Boltzmann equations and compare them to the numerical data in Sec.~\ref{subsec:boltzmann}.
In the analytical calculations, we assume periodic boundary conditions.

%%%%%%%%%%%%%%%%%%%%%%%%%%%%%
\subsection{Short-time dynamics} \label{subsec:shorttime}

The time dependence of an operator $O$ in the Heisenberg picture can be obtained for short times by using the Baker-Campbell-Hausdorff (BCH) formula
\begin{align} \label{bch}
\text{e}^{X} O \text{e}^{-X} = O + [X,O] + \frac{1}{2!} [X,[X,O]] + \cdots,
\end{align}
where $X=iHt$, and the $n$-th order expansion yields results up to ${\cal O}(t^n)$.
For the expectation value of the phonon-energy operator, one gets
\begin{align} \label{ephshorttime}
E_{\rm ph}(t) & =  \left \langle H_{\rm ph} \right \rangle -  \omega_0  \left \langle  H_{\rm coup}^{(1)} \right \rangle t \\ \nonumber
& + \left(  \omega_0 \gamma^2 
+ \frac{\omega_0^2}{2} \left  \langle  H_{\rm coup} \right \rangle
 - \frac{\omega_0}{2} \left  \langle  H_{\rm coup}^{(2)} \right \rangle \right) t^2 \\ \nonumber
& + {\cal O}(t^3),
\end{align}
where $\langle \dots \rangle$ denotes expectation values in the initial state.
Here, we have introduced expectation values of the generalized electron-phonon coupling term
\begin{equation}
H_{\rm coup}^{(\vartheta)} = - \frac{\gamma}{\sqrt{L}} \sum_{q,k} \left( M^{(\vartheta)}_{k,q} b_q^\dag c_{k-q}^\dag c_k + {\rm h.c.}  \right),
\end{equation}
where $M^{(1)}_{k,q} = i$ and $M^{(2)}_{k,q} = \epsilon_k - \epsilon_{k-q}$.
Even though $H_{\rm coup}^{(1)}$ and $H_{\rm coup}^{(2)}$ do not appear in the Holstein Hamiltonian~(\ref{holstein_ham}), which governs the time evolution, their expectation values may become relevant for the short-time dynamics.
For our initial state introduced in Eq.~(\ref{psi0}), however, most of the expectation values in Eq.~(\ref{ephshorttime}) vanish and we get 
\begin{equation} \label{eph_short}
E_{\rm ph}(t) = \omega_0 (\gamma t)^2 + {\cal O}(t^3).
\end{equation}

Similar derivations can be carried out for the other terms in the Hamiltonian.
For the kinetic energy, one gets
\begin{equation} \label{ekin_short}
E_{\rm kin}(t) = E_{\rm kin}(0) \left[ 1 - (\gamma t)^2 \right] + {\cal O}(t^3),
\end{equation}
while for the coupling energy, the result is
\begin{equation} \label{ecoup_short}
E_{\rm coup}(t) =   \left[  E_{\rm kin}(0) - \omega_0 \right] (\gamma t)^2 + {\cal O}(t^3).
\end{equation}
This clearly conserves the total energy during the time evolution,
$E_{\rm{total}} = E_{\text{kin}}(0) = E_{\text{ph}}(t) + E_{\text{coup}}(t) + E_{\text{kin}}(t)$.

\begin{figure}[!t]
\includegraphics[width=.96\columnwidth]{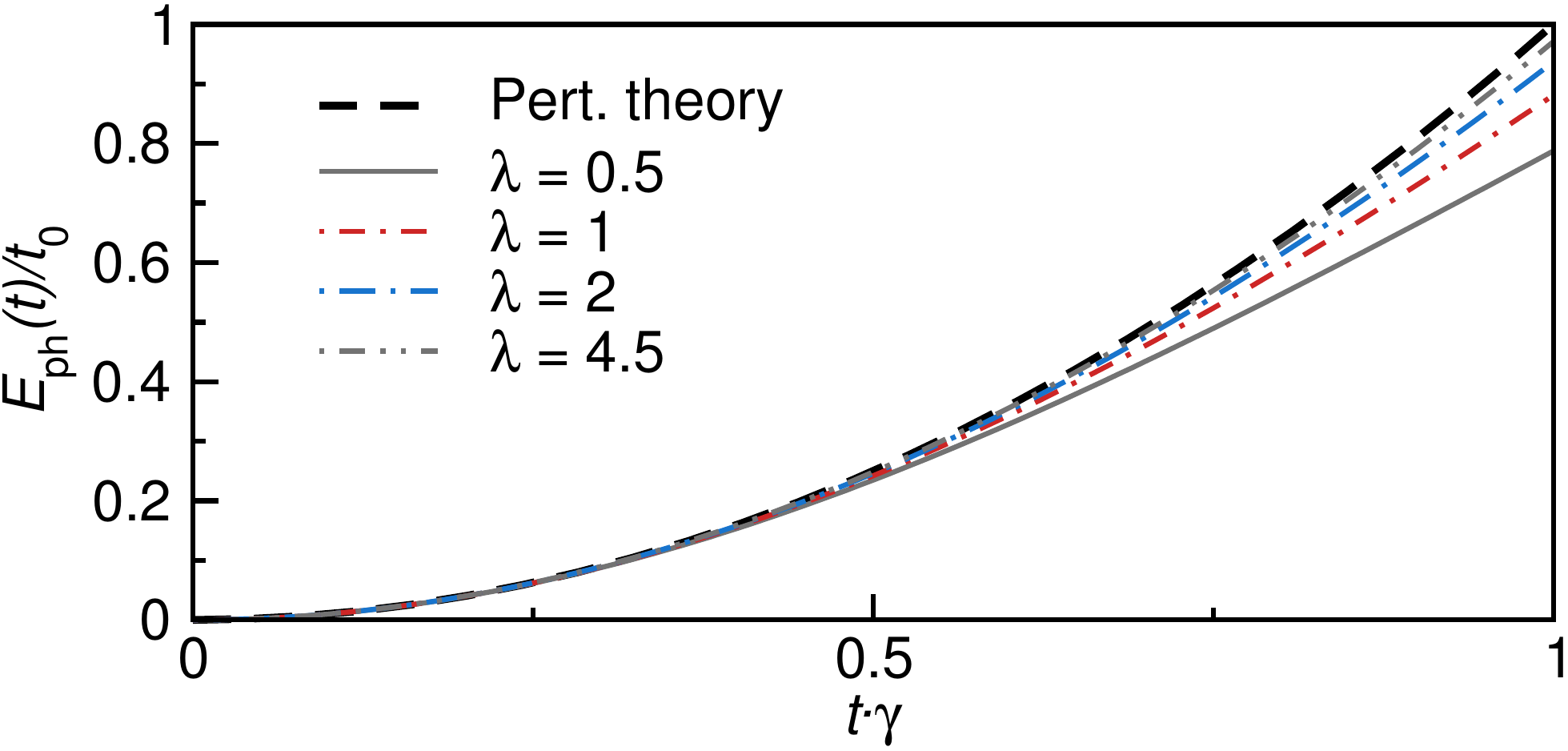}
\caption{(Color online)
Short-time dynamics of the phonon energy $E_{\rm ph}(t)$.
The bold-dashed line represents the results from perturbation theory, given by Eq.~(\ref{eph_short}).
All other curves are numerical results using LFS at $\omega_0 = t_0$ and different $\lambda$ (we use $L=12$ for $\lambda=0.5$ and $1$, and $L=8$ for $\lambda=2$ and $4.5$).
Time is measured in units of $1/\gamma$.
}
\label{fig:shorttime}
\end{figure}

For the initial state considered in our work, the short-time dynamics is therefore controlled by $1/\gamma$.
In Fig.~\ref{fig:shorttime} we compare the short-time evolution of $E_{\rm ph}(t)$ from Eq.~(\ref{eph_short}) with the numerical results using LFS.
The figure shows that the short-time expansion describes the numerical data well for $\gamma  t\lesssim 0.5$, while the agreement is the better the larger $\lambda$.

%%%%%%%%%%%%%%%%%%%%%%%%%%%%%%%%%%%%%%%%
\subsection{Perturbation theory in the interaction picture} \label{subsec:intpicture}

If one model parameter (for the sake of generality called $\eta$) is much smaller than other ones, we split the Hamiltonian into two terms
\begin{equation}
H = H_0 + \eta V.
\end{equation}
At short times, $\eta t$ represents a small parameter in the argument of the time-evolution operator $U(t)=\exp{(-iHt)}$.
It is convenient to expand $U(t)$ in the interaction picture
\begin{equation}
U(t) = U_0(t) + \eta U_1(t) + \eta^2 U_2(t) + {\cal O}(\eta^3),
\end{equation}
where
\begin{eqnarray}
U_0(t) & = & e^{-iH_0 t}, \\
U_1(t) & = & -i \int_{0}^t dt_1 e^{-iH_0 t_1} V e^{-iH_0 (t-t_1)}
\end{eqnarray}
and
\begin{eqnarray}
U_2(t) & = & - \int_{0}^t dt_1 \int_{0}^{t-t_1} dt_2 e^{-iH_0 t_1} V e^{-iH_0 t_2} \times \\ \nonumber
& & V e^{-iH_0 (t-t_1-t_2)}.
\end{eqnarray}
Hence, the time-dependent expectation values can be expressed in the initial
state as [we omit the time variable in $U_i(t)$]
\begin{eqnarray}
O(t)& =&  \left \langle U_0^\dagger O U_0 \right  \rangle \label{o_interacting}
+ \eta \left\langle U_0^\dagger O U_1+ U_1^\dagger O U_0 \right\rangle \\ \nonumber
& + &  \eta^2 \left\langle U_1^\dagger O U_1 + U_0^\dagger O U_2 + U_2^\dagger O U_0 \right\rangle \\ \nonumber
& + & {\cal O}(\eta^3),
\end{eqnarray}
which in many cases leads to accurate results in a longer time interval compared to the short-time expansion performed in Sec.~\ref{subsec:shorttime}.

%%%%%%%%%%%%%%%%%%%%%%%%%%%%
\subsection{Time evolution for weak electron-phonon coupling} \label{subsec:weakeph}

Here we explicitly derive $O(t)$ from Eq.~(\ref{o_interacting}) for the case $\eta=\gamma \ll \omega_0, t_0$ and our initial state~(\ref{psi0}).
We perform the derivation up to ${\cal O}(\gamma^2)$.
We then distinguish between the anti-adiabatic and adiabatic regime, i.e., Sec.~\ref{sssec:wanti} refers to the case $\omega_0 > 4t_0$ and Sec.~\ref{sss_relaxwc} to the case $\omega_0 < 4t_0$.

For the first-order contribution we get
\begin{multline} \label{eq_u0u1}
\left\langle U_0^\dagger O U_1+ U_1^\dagger O U_0 \right\rangle = \\ \frac{2}{\sqrt{L}} \sum_{q} R^{(1)}_{K,q} \left( \frac{ 1 - \cos{(\delta E_{K,q} t)} }{\delta E_{K,q} } \right),
\end{multline}
where
\begin{equation} \label{energy_cons}
\delta E_{K,q} = \epsilon_{K-q} + \omega_0 - \epsilon_K
\end{equation}
is the energy difference between one-phonon states (consisting of the electron 
with momentum $K-q$ and a phonon with momentum $q$) and the initial state (the electron with momentum $K$ and no phonon).
Electronic energies are given by the dispersion relation in Eq.~(\ref{e_dispersion}).
In addition,
\begin{equation}
R^{(1)}_{K,q}= \left \langle \emptyset \left \vert  c_K^{\phantom{\dagger}} O^{(\rm ele)} c_{K-q}^\dagger \right \vert \emptyset \right \rangle_{\text{ele}}
\left \langle \emptyset \left \vert  O^{(\rm ph)} b_{q}^\dagger \right \vert \emptyset \right \rangle_{\text{ph}}
\end{equation}
represents the matrix element decomposed into the electronic and the phononic part, and $O \equiv O^{(\rm ele)} \otimes O^{(\rm ph)}$.

The second-order contributions consist of two terms,
\begin{equation} \label{eq_u0u2}
\left\langle U_0^\dagger O U_2+ U_2^\dagger O U_0 \right\rangle = \\ -2 R^{(2a)}_K \left( \frac{ 1 - \cos{( \delta E_{K,q} t)} }{(\delta E_{K,q})^2} \right),
\end{equation}
and
\begin{equation} \label{eq_u1u1}
\left\langle U_1^\dagger O U_1 \right\rangle = \frac{2}{L} \sum_{q} R^{(2b)}_{K,q} \left( \frac{ 1 - \cos{(\delta E_{K,q} t)} }{(\delta E_{K,q})^2} \right).
\end{equation}
The corresponding matrix elements are
\begin{equation} \label{def_R2a}
R^{(2a)}_{K}= \left \langle \emptyset\left \vert  c_K^{\phantom{\dagger}}  O^{(\rm ele)} c_{K}^\dagger \right \vert \emptyset\right  \rangle_{\text{ele}}  
\left \langle \emptyset \left \vert  O^{(\text{ph})} \right \vert \emptyset \right \rangle_{\text{ph}}
\end{equation}
and
\begin{equation}
R^{(2b)}_{K,q,q'}= \left \langle \emptyset \left \vert  c_{K-q'}^{\phantom{\dagger}} O^{(\rm ele)} c_{K-q}^\dagger \right \vert \emptyset \right \rangle_{\text{ele}} 
  \left \langle \emptyset \left \vert  b_{q'}^{\phantom{\dagger}}  O^{(\rm ph)} b_{q}^\dagger \right \vert \emptyset \right \rangle_{\text{ph}},
\end{equation}
where the identity $R^{(2b)}_{K,q,q'} = R^{(2b)}_{K,q} \delta(q,q')$ has been
assumed to derive Eq.~(\ref{eq_u1u1}) (this is the case for all operators considered in this work),
and $O^{\rm (ph)}$ was assumed to be at most linear in $b$ in Eq.~(\ref{def_R2a}).

The expressions for the matrix elements $R^{(1)}_{K,q}$, $R^{(2a)}_{K}$ and $R^{(2b)}_{K,q}$ are listed in Appendix~\ref{sec:app1} for the observables of interest.
For the expectation values of the terms that appear in the Holstein
Hamiltonian~(\ref{holstein_ham}), we then get
\begin{eqnarray}
E_{\text{kin}}(t) & = & \epsilon_K \left( 1 - \frac{2\gamma^2}{L} \sum_q \frac{1 - \cos{ \left ( \delta E_{K,q} t \right )}}{\left (\delta E_{K,q}\right )^2 }  \right) \label{ekin_wc} \\
&& + \frac{2 \gamma^2}{L} \sum_{q} \epsilon_{K-q} \frac{1 - \cos{\left (\delta E_{K,q} t \right)} }{\left (\delta E_{K,q} \right)^2} \nonumber \\ 
E_{\text{ph}}(t) & = & \frac{2 \omega_0 \gamma^2}{L} \sum_{q} \frac{1 - \cos{\left (\delta E_{K,q} t \right)} }{\left (\delta E_{K,q}\right)^2}  \label{eph_wc} \\
E_{\text{coup}}(t) & = & - \frac{2 \gamma^2}{L} \sum_{q} \frac{1 - \cos{\left (\delta E_{K,q} t \right)}}{\delta E_{K,q}}. \label{ecoup_wc}
\end{eqnarray}

Note that the expansion of the above expressions up to $t^2$ matches the short-time results from Eqs.~(\ref{eph_short}),~(\ref{ekin_short}) and~(\ref{ecoup_short}).
Nevertheless, the results from Eqs.~(\ref{ekin_wc})-(\ref{ecoup_wc}) can be applied to a much larger time domain, as we are going to demonstrate in the following.

\subsubsection{Weak-coupling anti-adiabatic regime} \label{sssec:wanti}

\begin{figure}[!t]
\includegraphics[width=.96\columnwidth]{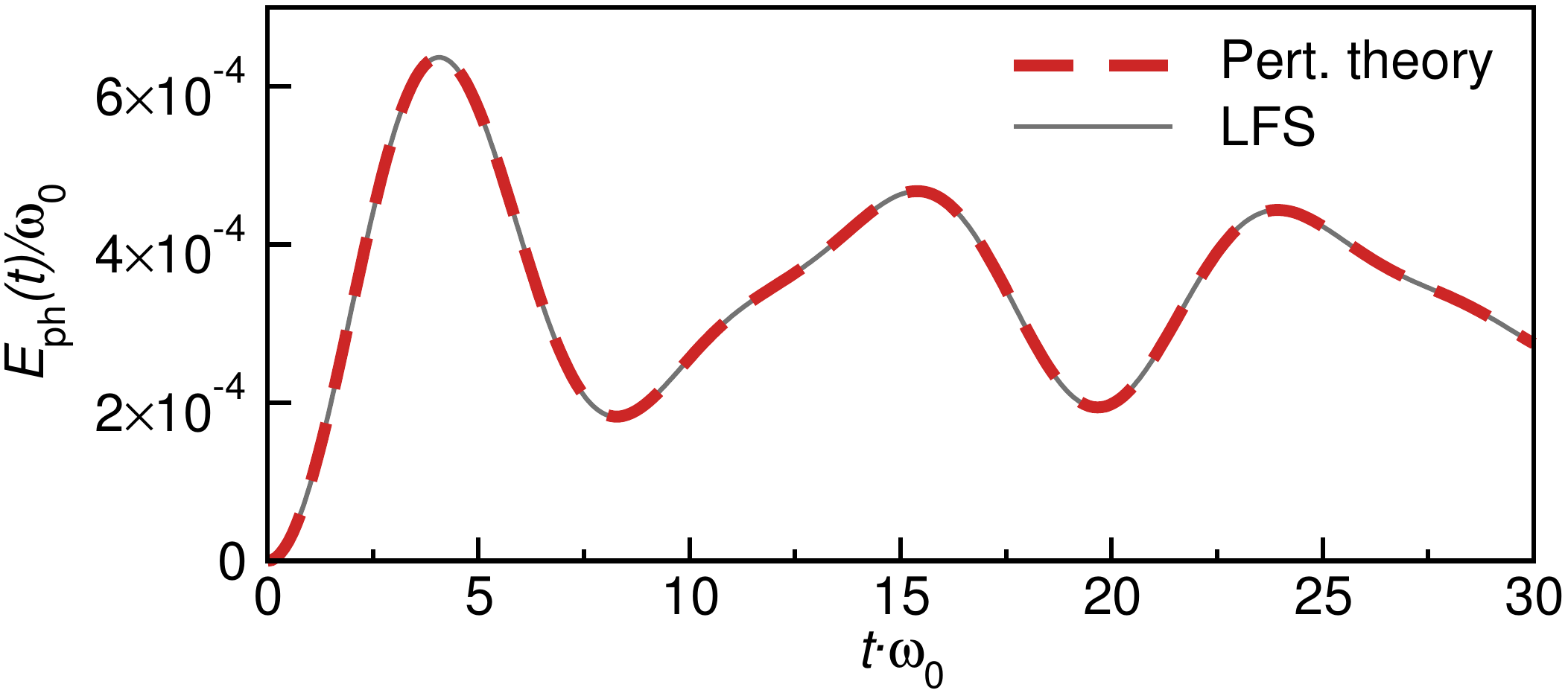}
\caption{(Color online)
Time evolution of the phonon energy $E_{\rm ph}(t)$ in the weak-coupling anti-adiabatic limit for $L=12$.
We set $t_0/\omega_0 = 0.1$ and $\gamma/\omega_0 = 0.01$, which corresponds to $\lambda=0.5 \times 10^{-3}$.
The solid line is the numerical result using LFS, while the dashed line is the result from perturbation theory, given by Eq.~(\ref{eph_wc}).
}
\label{fig:weak_antiadia}
\end{figure}

We first discuss the regime when the phonon energy is larger than the electronic bandwidth, $\omega_0 > 4t_0$
[more generally, if the initial $K \neq \pi$, it is sufficient to require $\omega_0 > 2t_0 (1-\cos{K})$].
Hence the energy of a single quantum phonon exceeds the maximal electronic energy difference, which already indicates the inefficiency of energy transfer.
To understand  the relevant energy scales in this regime, we first consider the anti-adiabatic limit $\omega_0 \gg t_0$ where
 we can replace $\delta E_{K,q}$ by   $\omega_0$ in Eqs.~(\ref{ekin_wc}), (\ref{eph_wc}) and~(\ref{ecoup_wc})
 and thus obtain simple formulas for the time evolution of these energies.
For the time-dependent phonon energy, this results in
\begin{equation} \label{eph_wc_antiadia}
E_{\rm ph}(t) = 2 \omega_0 \left( \frac{\gamma}{\omega_0} \right)^2  \left [ 1- \cos{(\omega_0t)} \right ] .
\end{equation}
There are two main observations from the above equation:
{\it (i)} The time evolution is governed by oscillations with frequency $\omega_0$;
{\it (ii)} The maximal number of emitted phonons  approaches $4 (\gamma/\omega_0)^2$.
This is qualitatively the same result as in the strong-coupling anti-adiabatic
regime ($\omega_0, \gamma \gg t_0$) to be discussed in Sec.~\ref{sssec:sclimit}, even though here we assume a weak coupling $\gamma \ll t_0$.

Nevertheless, when $1 < \omega_0 /(4t_0) \ll \infty$,
Eq.~(\ref{eph_wc_antiadia}) represents only a poor approximation, and one should
calculate the sum over $q$ in Eq.~(\ref{eph_wc}) explicitly.
In Fig.~\ref{fig:weak_antiadia}, we compare the resulting phonon energy 
with the numerically exact LFS data for $t_0 / \omega_0 = 0.1$ and $\gamma/\omega_0 = 0.01$.
The curves are virtually indistinguishable in the entire time interval $t \omega_0 < 30$.
Since Eq.~(\ref{eph_wc}) has been derived by taking into account transitions between the phonon vacuum state and  one-phonon states only, this implies that no higher-order processes are relevant for the dynamics in this parameter regime.

\subsubsection{Relaxation in the weak-coupling regime} \label{sss_relaxwc}

One of the central goals of our study is to investigate the relaxation dynamics, i.e., the situation where the majority of the electronic energy is, at sufficiently large times, transferred to phonons.
It is convenient to address this goal in the weak-coupling regime and for $\omega_0 < 2t_0 (1-\cos{K})$.
We study the electronic relaxation by measuring the temporal decrease of the electronic kinetic energy.
One can notice from Eq.~(\ref{ekin_wc}) that the kinetic energy is reduced with respect to the initial value by 
$\Delta E_{\rm kin}(t) = \sum_{q} \left ( \epsilon_{K-q} - \epsilon_{K} \right ) n_{K-q}(t)$,
where
\begin{equation} \label{def_nKq}
n_{K-q}(t) = \frac{4\gamma^2}{L} \frac{\sin^2{\left(\frac{\delta E_{K,q} t}{2}\right)}}{( \delta E_{K,q} )^2}.
\end{equation}
To address the relaxation, therefore, it is important to understand the time evolution of the momentum distribution function $n_k$.

We define the transition rate
$W = \frac{1}{\Delta t}\sum_{k \neq K} n_k(\Delta t)$,
which is the probability per time interval $\Delta t$ for the transition to electronic states different from the initial state.
In principle, this rate is time dependent, $W=W(\Delta t)$.

For our problem, following the notation of Eq.~(\ref{def_nKq}), it can be expressed as
\begin{equation}
W(\Delta t) = \frac{4\gamma^2}{L} \sum_{q\neq 0} \frac{1}{\Delta t} \frac{\sin^2{\left( \frac{\delta E_{K,q} \Delta t}{2} \right)}}{( \delta E_{K,q} )^2}.
\end{equation}
In the limit of large $L$, one can replace the sum by an integral and rewrite $W$ in the energy representation by introducing the dimensionless electronic density of states
\begin{equation}
{\cal D}_{\rm ele}(E  + \epsilon_0 - \omega_0) = 2t_0 \frac{{\rm d}q}{{\rm d}\epsilon_{K-q}} = \frac{1}{\sqrt{1 - \left( \frac{E  + \epsilon_0 - \omega_0}{2t_0} \right)^2}},
\end{equation}
where $E$ represents the electronic energy after the transition and $\epsilon_0$ is the electronic energy before the transition (i.e., if we start from the initial state, $\epsilon_0=\epsilon_K$).
The transition rate then equals
\begin{eqnarray} \label{transitionrate_3}
W(\Delta t) & = & \frac{\gamma^2}{t_0} \int_{-2t_0 - (\epsilon_0 - \omega_0)}^{2t_0 - (\epsilon_0 - \omega_0)} {\rm d} E \times \\ \nonumber
& & {\cal D}_{\rm ele}(E  + \epsilon_0 - \omega_0) \frac{2}{\pi} \frac{1}{\Delta t}  \frac{\sin^2{\left( \frac{E \Delta t}{2} \right)}}{E ^2}.
\end{eqnarray}
In the last part of the expression, one can recognize the delta function
$\delta (E) =2\sin{\left( \frac{E \Delta t}{2} \right)^2} /(\pi  E ^2 \Delta t)$
for sufficiently large time $\Delta t$.
Hence for short times, the electron can make transitions to various different energy levels with $\delta E_{K,q}\neq0$  , while for longer times, it transfers only to states with a well-defined energy satisfying the total energy conservation $\delta E_{K,q}=0$.
The time scale is given by the energy uncertainty principle $\Delta E \Delta t > \pi$.

Provided that $ {\cal D}_{\rm ele}(E  + \epsilon_0 - \omega_0) $ is a smooth function around $E=0$, this results in a time-independent transition rate
\begin{equation} \label{transitionrate_4}
W = \frac{\gamma^2}{t_0} {\cal D}_{\rm ele}(\epsilon_0 - \omega_0)  = 2 \lambda \omega_0 {\cal D}_{\rm ele}(\epsilon_0 - \omega_0).
\end{equation}
The transition rate is therefore proportional to the electronic density of states after the transition, known as Fermi's golden rule~\cite{ziman60}.
Equation~(\ref{transitionrate_4}) suggests expressing $W$ in units of $\omega_0$ since $\lambda$ and ${\cal D}_{\rm ele}$ are dimensionless by definition.
We will elaborate more on this issue in Sec.~\ref{sssec:const_density} where we show that it is also convenient to use the same unit to measure the characteristic time of the entire relaxation process.

In the adiabatic limit $\omega_0 \rightarrow 0$ the transition rate for an initial state with an arbitrary momentum $K$ is
\begin{equation}
W_K = \frac{\gamma^2}{t_0 \vert \sin(K) \vert}
\end{equation}
in leading order.
This result agrees up to a factor of two with the exact rate obtained in a 1D polaron model with a linear electronic dispersion~\cite{med96}.
The factor two is due to the presence of two electronic branches (and thus two relaxation paths) in the model~(\ref{holstein_ham}) as opposed to the single branch of the linear-dispersion model.
We can also understand why both results agree in the adiabatic limit only.
On the one hand, our model, Eq.~(\ref{holstein_ham}), and the linear-dispersion model of Ref.~\onlinecite{med96} are equivalent 
only if the transfer of electronic momentum $k$ is small.
On the other hand, Eq.~(\ref{def_nKq}) shows that the transition probability is maximal for $\delta E_{K,q}=0$.
Both conditions can be fulfilled simultaneously only in the adiabatic limit $\omega_0 \ll t_0$.

However, in 1D systems the electronic density of states diverges at the band edges and a more careful analysis 
is required for transitions to the bottom of the electronic band, i.e., for  $\epsilon_0 - \omega_0 \approx -2t_0$.
In that case,  Eq.~(\ref{transitionrate_3}) should be calculated explicitly.
By expanding ${\cal D}_{\rm ele}(E  -2 t_0) $ around $E=0$, we get
\begin{equation} \label{transitionrate_5}
W(\Delta t) = \frac{\gamma^2}{\sqrt{t_0}} \int_{0}^{4t_0} {\rm d} E
 \frac{2}{\pi} \frac{1}{\Delta t}  \frac{\sin^2{\left( \frac{E \Delta t}{2} \right)}}{E ^{5/2}}. 
\end{equation}
For $4t_0\Delta t \gg 1$ we obtain the asymptotic behavior
\begin{equation} \label{transitionrate_7}
W(\Delta t) \approx \frac{\gamma^2}{\sqrt{t_0}}  \frac{2\sqrt{2}}{3\sqrt \pi} \sqrt{\Delta t}.
\end{equation}
This shows that the transition rate to the lowest electronic state cannot be approximated by a time-independent quantity.

In the special case when the single phonon energy exactly matches the maximal electronic energy difference, $\omega_0 = 2t_0 (1-\cos{K})$, the result of Eq.~(\ref{transitionrate_7}) can be applied directly to get the time evolution of observables.
Consequently, we find an anomalous power-law behavior 
\begin{equation}
\Delta E_{\rm kin}(t) \approx - 2t_0 (1-\cos{K}) \sqrt{\frac{\gamma}{t_0}}  \frac{2\sqrt{2}}{3\sqrt \pi} (\gamma t)^{3/2}
\end{equation}
for times $t_0,\omega_0 \gg t^{-1} \gg \gamma$.

To summarize the discussion, the above equations were derived within the second-order perturbation theory describing the single-phonon emission only.
Hence, if $\omega_0 \ll 2t_0 (1-\cos{K})$, the two-level transition described by Eq.~(\ref{transitionrate_4}) is followed by a cascade of transitions forming the complete quasi-particle relaxation process (see also Fig.~\ref{fig:sketch}).
A possible way to describe this relaxation is to discretize time in small time steps and assume that at each step, the transition rates between the allowed electronic levels are given by Eq.~(\ref{transitionrate_4}) [or, in case of transitions to the bottom of the band, by Eq.~(\ref{transitionrate_7})].
We are going to pursue this idea in the context of the Boltzmann equation to calculate the time evolution of the electronic momentum distribution $n_k$.
Such dynamics is Markovian, but may still be a good approximation for weak enough electron-phonon couplings (for studies on the influence of Markovian effects on polaron dynamics see, e.g., Refs.~\cite{gelzinis11,dey14}).

\subsection{Boltzmann equation} \label{subsec:boltzmann}

The Boltzmann equation is a semi-classical approach for the non-equilibrium dynamics of electronic distribution functions in the thermodynamic limit.
In a system without any external force or inhomogeneity, the change of the momentum distribution comes solely from collisions.
The corresponding set of equations for the 1D electron-phonon system is
\begin{widetext}
\begin{eqnarray} \label{boltz1}
\dot{n}_{k} &=& \sum_q W_{k,q}
\left [
\left(
n_{{k-q}} (1-n_{k}) N_{q} 
 - n_{k} (1-n_{{k-q}})(N_{q}+1)
 \right) \right .
\delta\left(\epsilon_k - \epsilon_{k-q} - \hbar \omega_{q}\right) \\
&&
+ \left .\left(
n_{{k+q}} (1-n_{k})(N_{q}+1) - n_{k}(1-n_{{k+q}})N_q
\right)
\delta\left (\epsilon_k-\epsilon_{k+q}+\hbar \omega_q \right)
\right ]. \nonumber
\end{eqnarray}
\end{widetext}
The first half of the right-hand-side term describes transitions between the electronic state with momentum $k$ accompanied by $N_q$ phonons with momentum $q$ and the (lower-energy) electronic state with momentum $k$-$q$ and one more phonon.
The second half of the right-hand-side term describes transitions between the electronic state with momentum $k$ accompanied by $N_q+1$ phonons with momentum $q$ and the (higher-energy) electronic state with momentum $k$+$q$ and one less phonon.
The matrix element $W_{k,q}$ represents the transition rate for these processes.
In the following, we will rewrite Eq.~(\ref{boltz1}) in the energy representation $n_k \to n_\epsilon$ and use the transition rates from Eqs.~(\ref{transitionrate_4}) and~(\ref{transitionrate_7}).
In a 1D tight-binding model, there are two distinct $k$-points at the same energy (the exceptions being the top and the bottom of the band).
This was already taken into account in the derivation of the transition rates in Sec.~\ref{sss_relaxwc}, and the corresponding renormalization should be $(1-n_k) \to (1-n_\epsilon/2)$.
Consequently, the dynamics in energy representation consists of $s=4t_0/\omega_0$ transitions between $s+1$ electronic levels.
We assume, without loss of generality, that $s$ is an integer.

\subsubsection{Numerical solution}

We now solve the Boltzmann equation~(\ref{boltz1}) for the Holstein polaron
problem~(\ref{holstein_ham}) with our initial condition~(\ref{psi0}).
We set $\omega_q = \omega_0$ since our phonons are dispersionless.
Our initial state is a vacuum for phonons, and we therefore set $N_q=0$ at $t=0$ and assume the phonons to remain in that equilibrium state throughout the time evolution.
This implies that the electron never scatters again off phonons that it has
excited, which is a reasonable assumption when dealing with a single electron in a large empty lattice.
In total, there are $s+1$ equations for $s$ transitions, where in each transition, the electronic energy is lowered by $\omega_0$,
\begin{eqnarray} \label{boltz2}
\dot{n}_{\epsilon_0} &=& - W_1 \, n_{\epsilon_0} (1-n_{\epsilon_1}/2) \nonumber \\
\dot{n}_{\epsilon_1} &=& - W_2 \, n_{\epsilon_1} (1-n_{\epsilon_2}/2) + W_1 \, n_{\epsilon_0} (1-n_{\epsilon_1}/2) \nonumber \\
& \dots &  \\
\dot{n}_{\epsilon_{s-1}} &=& - W_{s} \, n_{\epsilon_{s-1}} (1-n_{\epsilon_{s}}) + W_{s-1} \, n_{\epsilon_{s-2}} (1-n_{\epsilon_{s-1}}/2) \nonumber \\
\dot{n}_{\epsilon_s} &=&  \hspace{0.27cm}  W_s \, n_{\epsilon_{s-1}} (1-n_{\epsilon_s}). \nonumber
\end{eqnarray}
The nearest energy levels are hence related by $\epsilon_{m+1} = \epsilon_{m}-\omega_0$.
For $1 \leq m < s$, the transition rate is $W_m = (\gamma^2/t_0) {\cal D}_{\rm ele}(\epsilon_m)$.
In the last equation, $W_s$ is time-dependent and given by Eq.~(\ref{transitionrate_7}).

\begin{figure}[!t]
\includegraphics[width=.96\columnwidth]{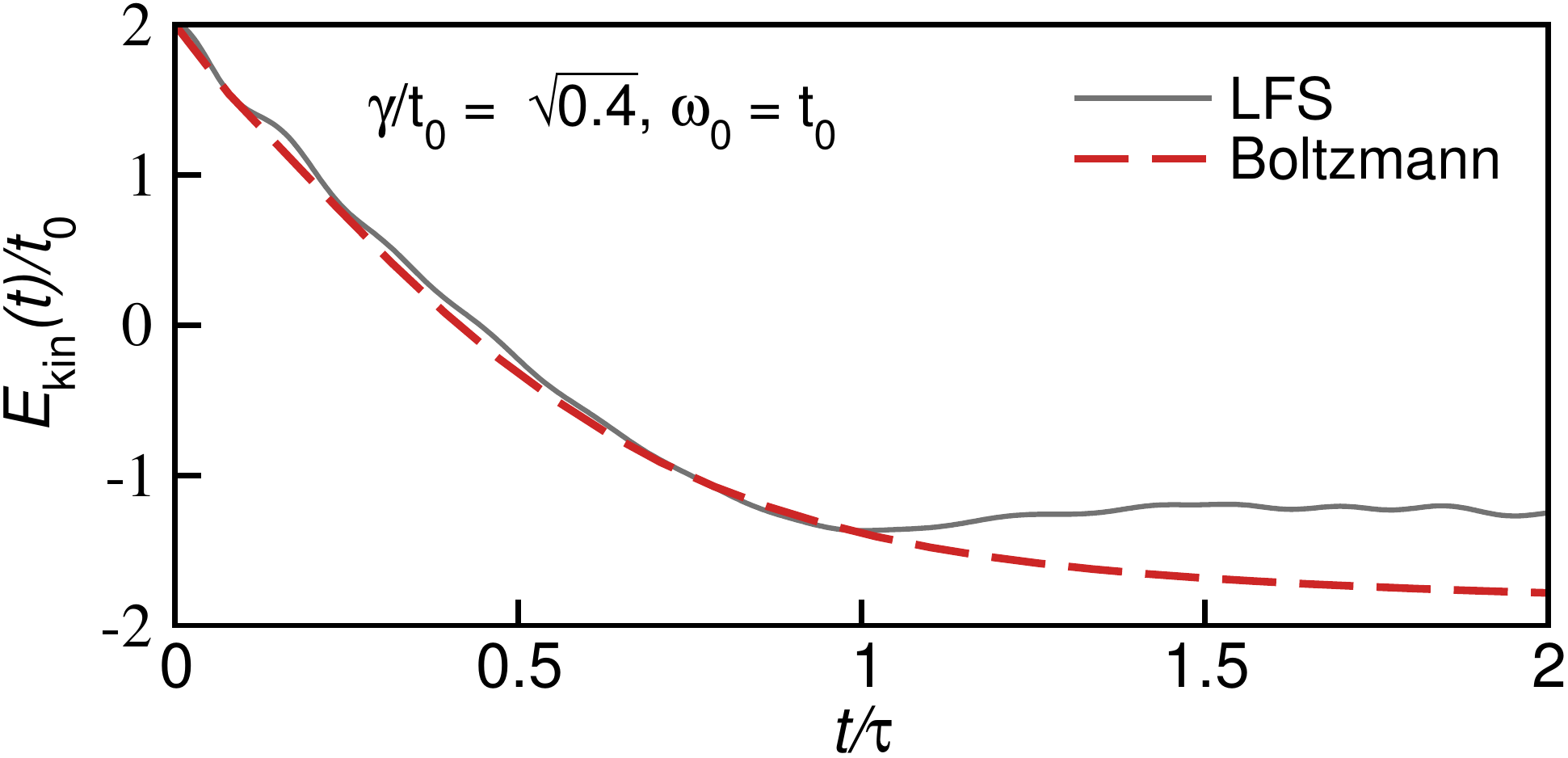}
\caption{(Color online)
Relaxation of the electronic kinetic energy $E_{\rm kin}(t)$ at $\omega_0=t_0$ in the weak-coupling regime $(\gamma/t_0)^2=0.4$, which corresponds to $\lambda=0.2$.
The solid line is the numerical result using LFS (for $L=12$), the dashed line is the result from the numerically integrated Boltzmann equations~(\ref{boltz2}).
Time is measured in units of $\tau$, defined in Eq.~(\ref{tau_constantdos}).
}
\label{fig:boltzmann}
\end{figure}

When solving Eq.~(\ref{boltz2}) numerically, some care is required when choosing the appropriate discrete time step.
On the one hand, in the derivation of the time-independent values of $W_m$ [see Eq.~(\ref{transitionrate_3})] we require that the time step is not too small;
on the other hand, the equations must necessarily fulfill $W_m \Delta t < 1$, which implies $\Delta t \omega_0 \alt  (2 \lambda)^{-1}$ (we assumed ${\cal D}_{\rm ele}\sim 1$ for simplicity).
For our calculation, we therefore took $\Delta t \omega_0 = 1$.
In Fig.~\ref{fig:boltzmann} we compare the electronic kinetic energy $E_{\rm kin}(t) = \sum_k \epsilon_k n_k(t)$ obtained by the Boltzmann kinetic equation, with the numerical solution using LFS.
Remarkably, the results show perfect agreement in the relaxation regime.
This indicates that, as long as the electron-phonon coupling is sufficiently small, the processes described by Eq.~(\ref{boltz2}) provide the most relevant contribution to the dynamics.
Deviations set in when the steady state is approached: there, the dynamics on a
finite lattice is governed by the interplay between the electron and the phonons
emitted during the relaxation.
Clearly, such processes are not included in the Boltzmann equations~(\ref{boltz2}), and thus an agreement between the solution of the Holstein model and the Boltzmann equation is not expected in the steady state.
We mention that recently, by using the Boltzmann approach, the mobility of the Holstein polaron was accurately reproduced in the weak-coupling regime~\cite{mishchenko14}.

\subsubsection{Relaxation for a constant density of states} \label{sssec:const_density}

\begin{figure}[!t]
\includegraphics[width=.96\columnwidth]{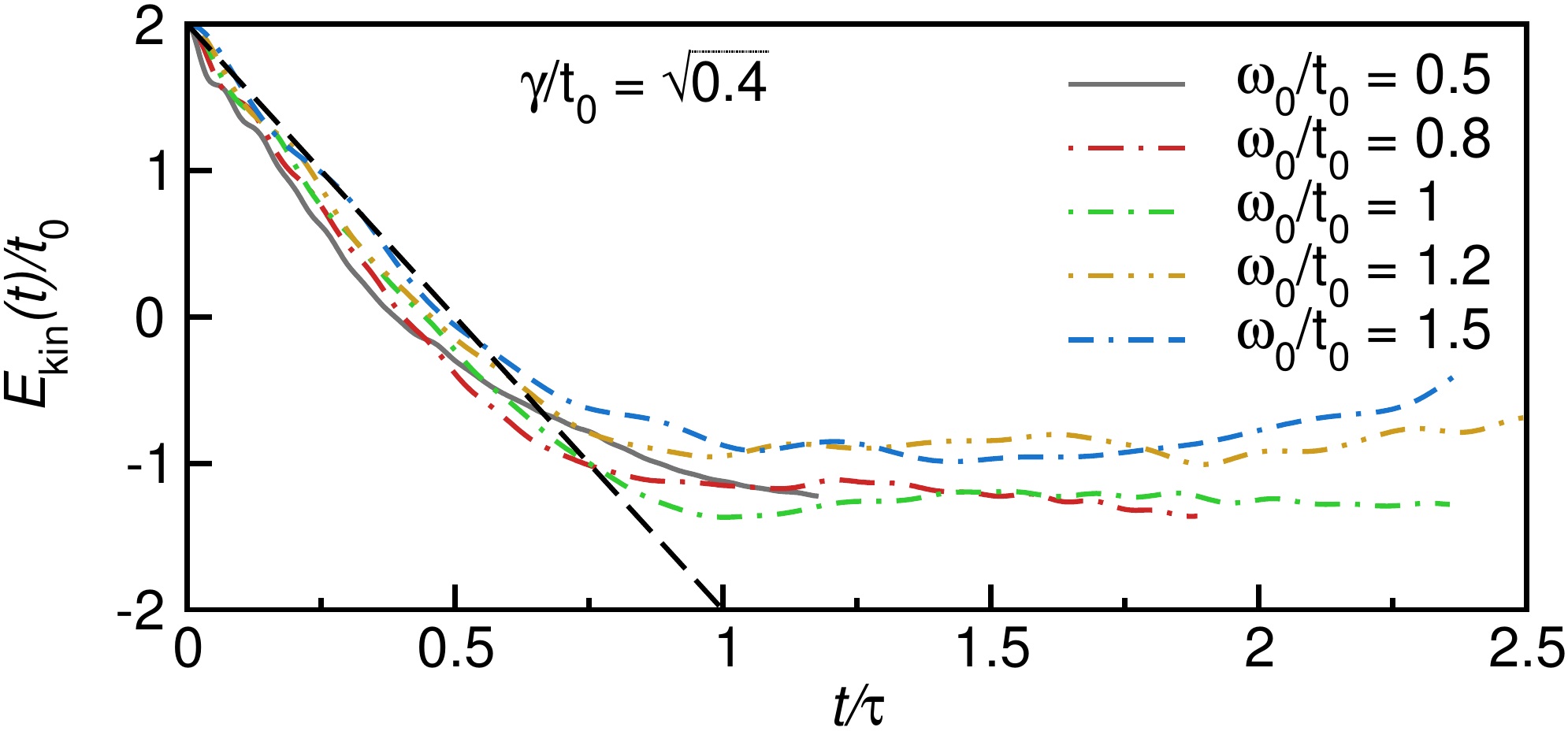}
\caption{(Color online)
Relaxation of the electronic kinetic energy $E_{\rm kin}(t)$ in the weak-coupling regime.
Results are shown for a fixed $\gamma$ and $t_0$ [setting $(\gamma/t_0)^2=0.4$] and different values of $\omega_0/t_0 =0.5,0.8,1.0,1.2,1.5$, which correspond to $\lambda=0.4,0.25,0.2,0.17,0.13$, respectively.
We use LFS for $L=12$.
The dashed line is the function $E_{\rm kin}(t)/t_0=2[1-2(t/\tau)]$ with the relaxation time $\tau$ defined in Eq.~(\ref{tau_constantdos}).
}
\label{fig:taulfs}
\end{figure}

The occupation of different electronic $k$-states can also be obtained analytically under some approximations.
We can linearize the Boltzmann equations~(\ref{boltz2}) because we have only one
electron in the system and we can assume a constant energy density ${\cal D}_{\rm ele} = \pi/4$, 
which gives the energy-independent transition rate $\Gamma = (\pi/2)\lambda \omega_0$
(this would be exact for a linear electronic dispersion~\cite{med96}).
Then one needs to solve equations of the form
\begin{eqnarray}
\dot{n}_{\epsilon_0} &=& - \Gamma \, n_{\epsilon_0} \nonumber \\
\dot{n}_{\epsilon_1} &=& - \Gamma \, n_{\epsilon_1} + \Gamma \, n_{\epsilon_0} \nonumber \\
& \dots & \\
\dot{n}_{\epsilon_{s-1}} &=& - \Gamma \, n_{\epsilon_{s-1}} + \Gamma \, n_{\epsilon_{s-2}} \nonumber \\
\dot{n}_{\epsilon_s} &=& \hspace{0.27cm}  \Gamma \, n_{\epsilon_{s-1}}. \nonumber
\end{eqnarray}
The solution is, for a fixed $0\leq m < s$,
\begin{equation} \label{nt_poisson}
n_{\epsilon_m}(t) = \frac{(\Gamma t)^m}{m!} e^{-\Gamma t},
\end{equation}
which is a Poisson distribution and has a maximum at $m\approx \Gamma t $.
To obtain the characteristic relaxation time $\tau$, we simply require $\tau \equiv s/\Gamma$.
This definition provides a reasonable estimate for the time when the lowest single-electron state (i.e., the one at $k=0$) becomes dominantly occupied.
It gives
\begin{equation} \label{tau_constantdos}
\tau \omega_0 = \frac{16}{\pi} \left( \frac{\gamma}{t_0}\right)^{-2} = \frac{8}{\pi} \left(\lambda \frac{\omega_0}{t_0} \right)^{-1}.
\end{equation}
The main observation from this result is that in general, the quantitative value of the characteristic relaxation time depends on all the three parameters of the Holstein Hamiltonian.
Nevertheless, an important insight from Eq.~(\ref{tau_constantdos}) is that $1/\omega_0$ is a very convenient time unit to measure relaxation.
The advantage of measuring time as $t \omega_0$ is twofold:
first, since the derivation stems from the weak-coupling regime, $\tau \omega_0$ is always larger than one,
and second, the quantitative value of the relaxation time in these units is then only a function of the electron-phonon coupling energy in units of the electron delocalization energy.
In addition, the result~(\ref{tau_constantdos}) also affirms that the relaxation time can not be deduced from the short-time expansion in Eq.~(\ref{ekin_short}).

We test the prediction of the relaxation time~(\ref{tau_constantdos}) using the numerical results for the Holstein Hamiltonian.
In Fig.~\ref{fig:taulfs} we plot $E_{\rm kin}(t)$ for different $\omega_0$, but for fixed values of $\gamma$ and $t_0$.
When time is measured in units of $\omega_0$, the curves fairly well collapse on the same line in the relaxation regime.
Remarkably, the simple estimate~(\ref{tau_constantdos}), using a constant density of states, provides a reasonable measure for the relaxation time.
Physically, the reason for this agreement is that the relaxation is dominated by the slowest transitions $W_m$ in the Boltzmann equations~(\ref{boltz2}), i.e., when the electron is away from the band edges and their divergent density of states.
Hence even though both initial and final electronic states are at the band edges, a decent estimate for the relaxation time can already be obtained from the assumption of a constant density of states.

%%%%%%%%%%%%%%%%%%%%%%%%%%%%
\subsection{Time evolution for small hopping amplitude} \label{sssec:sclimit}

We now consider the case $t_0 \ll \gamma, \omega_0$.
This is the limit where the net energy transfer between the electron and phonons is expected to be small or even does not occur.
Instead, the time dependence of observables exhibits strong oscillations.
Analytical solutions for the lowest non-zero orders in $\eta=t_0$ allow us to show explicitly that the period of oscillations is given by 
$\omega_0$, while the amplitude of oscillation is governed by the ratio $g=\gamma/\omega_0$.

\subsubsection{Single-site dynamics} \label{sssec:t0zero}

For $t_0=0$ the unperturbed Hamiltonian $H_0$ is given by Eq.~(\ref{sec2:effham}).
As all lattice sites are decoupled, it can be diagonalized 
by means of a rotation to the coherent-state basis~\cite{langfirsov} as shown in
Sec.~\ref{sec:setup}.
Therefore, the exact analytical solution can be obtained for all times.
To carry out a perturbative expansion in $t_0$, however, it is more convenient to rotate
the Hamiltonian operator instead of the basis. Explicitly, any operator $O$ can be rotated
by a unitary Lang-Firsov transformation
\begin{equation} \label{operator_trans}
\widetilde{O} = e^{iS} O e^{-iS}
\end{equation}
with the hermitian operator
\begin{equation}
S = g \sum_j p_j n_j,
\end{equation}
where $p_j = i (b_j^\dagger - b_j^{\phantom{\dagger}})$ represents the local
oscillator momentum operator.
The Hamiltonian is then diagonal
\begin{equation}
\widetilde{H}_0 = \omega_0 \sum_j b_j^\dagger b_j^{\phantom{\dagger}} - \varepsilon_b
\end{equation}
with the polaron binding energy $\varepsilon_b$ defined
in Eq.~(\ref{polaron_energy}) and the initial state~(\ref{psi0}) is now represented
 by 
\begin{eqnarray}
\ket{\widetilde{\psi}_0} & = & e^{iS} \ket{\psi_0} \\
& = & 
\frac{e^{-g^2/2}}{\sqrt{L}} \sum_j e^{ij K}
\left [ e^{-g b_j^\dagger} \ket{\emptyset}_{\rm ph}  
\otimes c_j^{\dag} \ket{\emptyset}_{\rm ele} \right ].  \nonumber
\end{eqnarray}
The time evolution of this state can be calculated exactly
\begin{eqnarray}
\label{eq:solution2}
\ket{\widetilde{\psi}(t)} & = & e^{-i\widetilde{H}t} \ket{\widetilde{\psi}_0}  \\
& = & e^{i\varepsilon_b t}
\frac{e^{-g^2/2}}{\sqrt{L}} \sum_j e^{ijK}
\left [ e^{-g(t) b_j^\dagger} \ket{\emptyset}_{\rm ph}  
\otimes c_j^{\dag} \ket{\emptyset}_{\rm ele} \right ] \nonumber
\end{eqnarray}
with $g(t) = g e^{-i\omega_0t}$.
Clearly, this state describes a delocalized composite quasi-particle made of
the electron dressed by a phonon cloud. This is qualitatively similar 
to the small polaron found in the ground state of the Holstein model.
However, here the phonon cloud fluctuates with time in 
contrast to the static cloud of the ground-state polaron.

Finally, time-dependent expectation values can be calculated directly in this
representation using
\begin{equation} \label{o_coherent}
O(t) = \bra{\widetilde{\psi}(t)} \widetilde{O} \ket{\widetilde{\psi}(t)}.
\end{equation}
Using the BCH formula~(\ref{bch}) we easily get 
\begin{equation}
\widetilde{x}_j = b_j^\dagger + b_j^{\phantom{\dagger}} + 2g n_j
\end {equation}
for the local oscillator displacement operator $x_j = b_j^\dagger + b_j^{\phantom{\dagger}}$ 
and $\widetilde{p}_j = p_j$ for the corresponding momentum operator. 
Thus their expectation values are 
\begin{eqnarray}
x_j(t) & = & \frac{2 g}{L} \left[ 1- \cos{(\omega_0 t)} \right] 
\\
p_j(t) & = & \frac{2 g}{L} \sin{(\omega_0 t)}.
\end{eqnarray}
This result can be easily understood. Each site has a probability $1/L$ to be
occupied by the electron. If the site is occupied, the boson degree of freedom
represents a harmonic oscillator with frequency $\omega_0$ and an equilibrium 
position $\langle x \rangle_{\text{eq}} = 2g$.
Our initial condition~(\ref{psi0}) corresponds to $x_j(t=0)=p_j(t=0)=0$.
Thus, the oscillator swings between $x=0$ and $x=4g$ like a classical harmonic
oscillator. 
Similarly, for the phonon energy $H_{\text{ph}} = \omega_0 \sum_j  b_j^\dagger b_j^{\phantom{\dagger}}$ we obtain 
\begin{equation} 
\widetilde{H}_{\text{ph}} =  \omega_0 \sum_j  b_j^\dagger b_j^{\phantom{\dagger}}
+ g \omega_0 \sum_j  \left (b_j^\dagger +b_j^{\phantom{\dagger}} \right) n_j 
+ \varepsilon_b
\end{equation} 
and then the expectation value 
\begin{eqnarray}
E_{\text{ph}}(t) & = & 2 g^2 \omega_0 \left[ 1- \cos{(\omega_0 t)} \right]
\label{nph_t0limit}. 
\end{eqnarray}
This is identical to the result~(\ref{eph_wc_antiadia}) obtained in the
anti-adiabatic limit of the weak-coupling perturbation expansion, i.e., for $\gamma \ll t_0 \ll \omega_0$, although we have assumed 
$t_0 \ll \gamma,\omega_0$ to derive the present result.
Thus in the anti-adiabatic limit, Eq.~(\ref{nph_t0limit}) seems to hold in both weak- and strong-coupling regime.
This is also confirmed by our TEBD simulations for $t_0/\omega_0 = 10^{-3}$ and finite $\gamma$.
We also note that the number of phonons $N_{\text{ph}}(t)=E_{\text{ph}}(t)/\omega_0$ reaches
a maximum value $4 g^2$ as a function of time, i.e., four times larger
than the polaron ground-state value $N_{\rm ph} = g^2$ in the anti-adiabatic
strong-coupling regime.
In all observables of interest, the oscillations never decay in time and their
amplitude is determined by the bare electron-phonon coupling $\gamma$ expressed in units of $\omega_0$. 
All results also unambiguously show that $2\pi/\omega_0$ or integer fractions
thereof set the period of these oscillations.
The time evolution therefore imposes harmonic oscillations of the system 
 around its polaron ground state.

\subsubsection{Finite but small hopping amplitude} \label{sssec:t0small}

\begin{figure}[!t]
\includegraphics[width=.96\columnwidth]{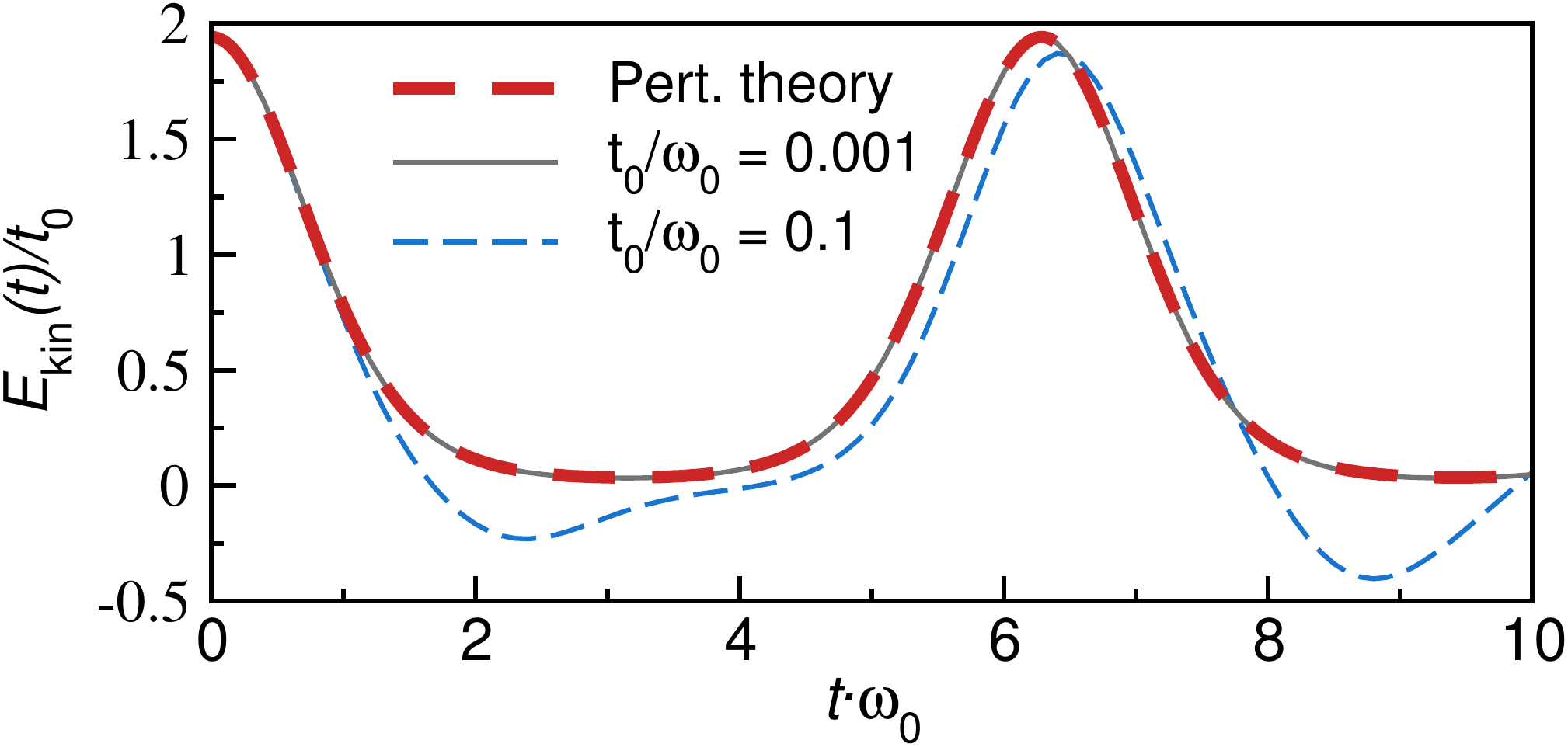}
\caption{(Color online)
The electronic kinetic energy $E_{\rm kin}(t)$ in the strong-coupling anti-adiabatic regime at $\gamma/\omega_0 = 1$ and $L=12$.
The dashed line is the result from perturbation theory, given by Eq.~(\ref{eq:Ekin_antiad}),
while the solid and thin dashed line represent the TEBD result for $t_0/\omega_0 = 10^{-3}$ and $10^{-1}$ (corresponding to $\lambda=500$ and $5$, respectively).
}
\label{fig:E_kin_antiad}
\end{figure}

If the hopping amplitude is finite but still $t_0 t \ll 1$, the above results can be seen as the zeroth-order term
of an expansion~(\ref{o_interacting}) in power of $\eta=t_0 \ll t^{-1}$.
The time evolution of the first-order term of the kinetic energy 
\begin{equation}
E_{\text{kin}}(t) = \sum_k \epsilon_k n_k(t)
\end{equation}
can also be obtained from zeroth-order expectation values~(\ref{o_coherent}).
Using 
\begin{equation}
\widetilde{c}^{\dagger}_k \widetilde{c_k} =  \frac{1}{L} \sum_{j,l} e^{ik(j-l)} e^{ig(p_j-p_l)} c_j^{\dag} c_l^{\phantom{\dag}} 
\end{equation}
we first calculate the electronic momentum distribution in zeroth order resulting in
\begin{equation}
n_k(t) = \frac{1}{L} + e^{2 g^2 \left [\cos{(\omega_0 t ) - 1} \right]} \left ( \delta_{k,K} -
\frac{1}{L} \right ) 
\end{equation}
and then obtain 
\begin{equation}
E_{\text{kin}}(t) = -2 t_0 \cos(K) e^{2 g^2 \left [\cos(\omega_0 t ) - 1 \right]} . \label{eq:Ekin_antiad}
\end{equation}
This result is plotted in Fig.~\ref{fig:E_kin_antiad}.
We see that the kinetic energy is also a periodic function with period $2\pi/\omega_0$ that oscillates between its initial value $2t_0\cos(K)$ and an exponentially reduced value $\exp(-4 g^2) 2t_0\cos(K)$. 
This confirms that the system oscillates without any relaxation up to leading orders in $t_0$ in the limit of small hopping terms $t_0$.
Since the result~(\ref{eq:Ekin_antiad}) is valid for both open and periodic boundary conditions, we can compare it
directly to TEBD simulations using an initial standing wave with the wave number $K=\pi L/(L+1)$.
We observe a perfect agreement between Eq.~(\ref{eq:Ekin_antiad}) and the TEBD result for $t_0/\omega_0 = 10^{-3}$, as shown in Fig.~\ref{fig:E_kin_antiad}.
We note that the kinetic energy scale is a tiny fraction of the other energy scales, i.e., $E_{\text{kin}}(t) \alt 10^{-3} \omega_0, \varepsilon_b$, but it is nevertheless perfectly reproduced by the TEBD data.
This shows that TEBD simulations are very accurate in this regime.
When the ratio $t_0/\omega_0$ increases, the numerical results and perturbation theory cease to perfectly agree with each other.
However, we show in Fig.~\ref{fig:E_kin_antiad} that at $t_0/\omega_0 = 0.1$, Eq.~(\ref{eq:Ekin_antiad}) still provides the correct qualitative behavior of the dynamics.

%%%%%%%%%%%%%%%%%%%%%%%%%%%%%%%%%%%%%%%%%%%%%%%%%%%%%%%%%%%%%%%%%%%%%%%%%%%%%%%
\section{Numerical results for intermediate parameter regimes} \label{sec:results}

In this section we focus on non-perturbative results by applying diagonalization in the LFS.
This complements Sec.~\ref{sec:perturbative} that has addressed the regimes where one parameter of the Hamiltonian is vanishingly small.
The central question here is to which degree the properties of the weak- and strong-coupling limits as well as adiabatic and anti-adiabatic limits persist at intermediate values of parameters.

%%%%%%%%%%%%%%%%%%%%%%%%%%%%
\subsection{Crossover from adiabatic to anti-adiabatic regime} \label{sec:adiaantiadia}

\begin{figure}[!t]
\includegraphics[width=.96\columnwidth]{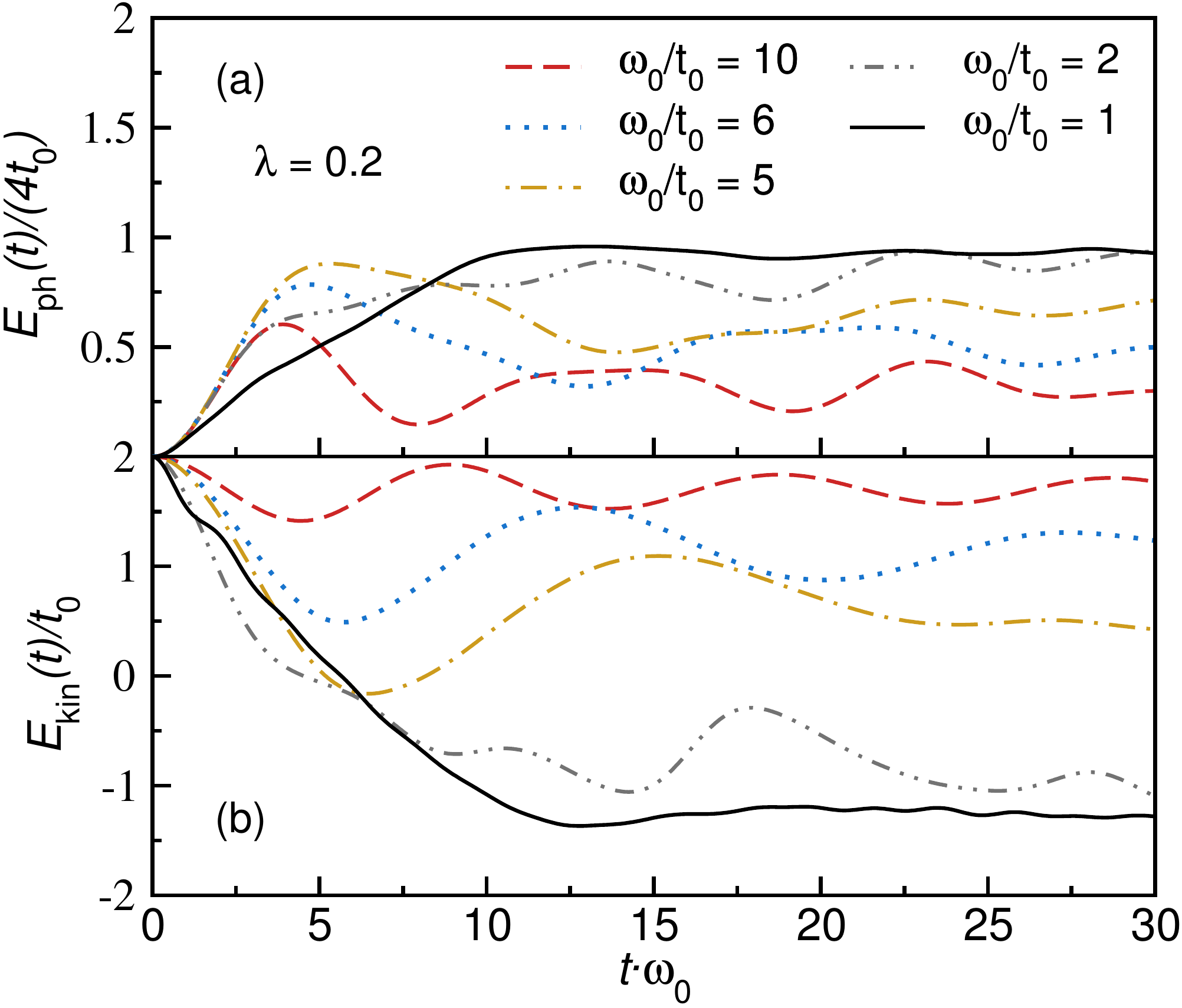}
\caption{(Color online)
Time evolution for the two parts of the Hamiltonian:
(a) the phonon energy divided by the bandwidth $E_{\rm ph}(t)/(4t_0)$ and (b) the electron kinetic energy $E_{\rm kin}(t)$, plotted for several values of the adiabaticity ratio $\omega_0/t_0$.
We use LFS at $\lambda=0.2$ and $L=12$.
}
\label{fig:Eph_lambda}
\end{figure}

The solution of the second-order perturbation theory in the weak-coupling anti-adiabatic regime discussed in Sec.~\ref{sssec:wanti} yields oscillatory behavior of all parts of the Hamiltonian
[see Fig.~\ref{fig:weak_antiadia} for $E_{\rm ph}(t)$].
%In Fig.~\ref{fig:weak_antiadia} we show the numerical result for $E_{\rm ph}(t)$, which perfectly fits to the perturbative expansion in Eq.~(\ref{eph_wc}).
In this case relaxation barely takes place since the net energy transfer from electron to phonons is negligible and comparable to the amplitude of oscillations.
On the other hand, the adiabatic regime at weak electron-phonon coupling is reasonably well described by the Boltzmann equation, see Sec.~\ref{subsec:boltzmann}.
This is the paradigmatic case for relaxation since the majority of the electronic kinetic energy is transferred to phonons.
%We showed in Fig.~\ref{fig:taulfs} that the Boltzmann equations provide a reasonable description of relaxation dynamics up to $\omega_0/t_0 \lesssim 2$.
The question remains how these two regimes evolve into each other when the ratio $\omega_0/t_0$ is varied.

A convenient way of addressing this is to study the weak-coupling regime where $\gamma$ is smaller than $\omega_0$ and $4t_0$. 
In this case, the amount of emitted phonons in the adiabatic regime roughly compensates the reduction of the electronic kinetic energy.
In Figs.~\ref{fig:Eph_lambda}(a) and~\ref{fig:Eph_lambda}(b) we plot, for $\lambda=0.20$, the phonon energy divided by the bandwidth $E_{\rm ph}(t)/(4t_0)$ and the electron kinetic energy $E_{\rm kin}(t)$, respectively.
At $\omega_0 = t_0$ we observe $E_{\rm ph}(t)/(4t_0) \approx 1$ at sufficiently long times and a considerable reduction of the electronic kinetic energy.
On the other hand, when $\omega_0 \gg t_0$, $E_{\rm ph}(t)/(4t_0) < 1$ for all times, and the majority of the excess energy remains in the electronic sector.
Our results in Fig.~\ref{fig:Eph_lambda} show that there is a continuous crossover from the regime at $\omega_0 \ll 4t_0$, dominated by relaxation of the electronic kinetic energy, towards a different type of behavior at $\omega_0 > 4t_0$ governed by coherent oscillations and only weak redistribution of energies.

%%%%%%%%%%%%%%%%%%%%%%%%%%%%
\subsection{Crossover from weak to strong coupling at $\omega_0 = t_0$} \label{sec:crosslambda}

In the following we focus on the role of the electron-phonon coupling at $\omega_0 = t_0$ and study the dynamics at the crossover from weak to strong coupling.
We describe the features in two different time regimes, the relaxation regime and the stationary regime, in Secs.~\ref{subsec:relax} and~\ref{subsec:steady}, respectively.

\subsubsection{Relaxation regime} \label{subsec:relax}

\begin{figure}[!t]
\includegraphics[width=.99\columnwidth]{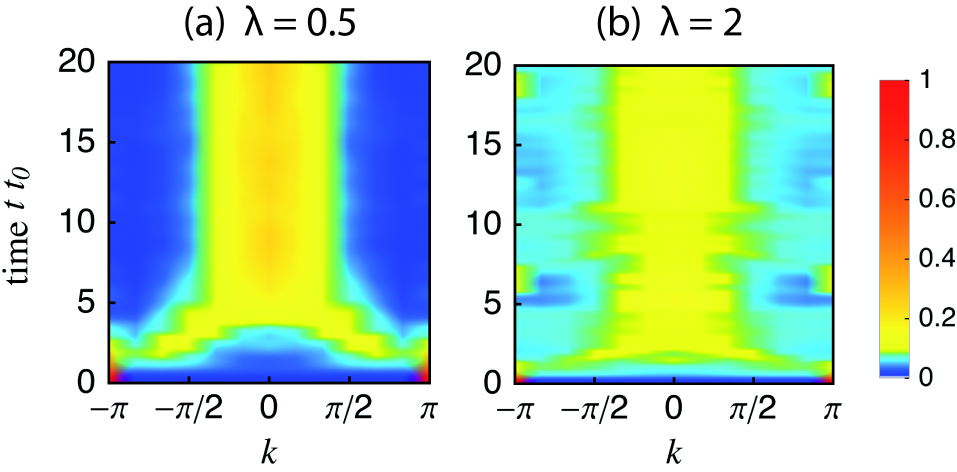}
\caption{(Color online)
Density plot showing the electronic momentum distribution function $n_k(t)$ using LFS at $\omega_0=t_0$ and $L=12$.
(a) $\lambda=0.5$, (b) $\lambda=2$.
}\label{fig:mdf}
\end{figure}

The relaxation regime is the regime in which excess energy is transferred from the electron to the phonon degrees of freedom.
This occurs in the first stage of the time evolution, i.e., in the time interval roughly given by the characteristic relaxation time.
In Sec.~\ref{sec:perturbative}, perturbative arguments allowed us to clearly distinguish between the limiting cases where relaxation is expected to take place and where it is inefficient: a sufficient condition for the first scenario is that the phonon energy is much smaller than the electronic bandwidth.
The decrease of kinetic energy is a consequence of the redistribution of the electronic momentum distribution, discussed in the context of the Boltzmann equation in Sec.~\ref{subsec:boltzmann}.
The simplest picture of momentum redistribution (relevant, in particular, in the weak-coupling regime) has already been discussed in the context of Fig.~\ref{fig:sketch}:
the electron reduces its kinetic energy by emitting phonons, where each of the processes conserves the total energy and momentum.
In Fig.~\ref{fig:mdf}(a) we show the evolution of the electronic momentum distribution function $n_k$ as a function of time at moderate coupling $\lambda=0.5$ and $\omega_0=t_0$.
The electron starts off with a momentum $k=\pi$ due to our initial condition~(\ref{psi0}).
The momentum redistribution can be seen at short times in the density plot of Fig.~\ref{fig:mdf}(a) as a narrow bright region expanding roughly linearly towards $k = 0$.
When the maximum occupation is at $k= 0$ [this occurs at $t t_0
\approx 5$ for $\lambda=0.5$ and $\omega_0 = t_0$ in Fig.~\ref{fig:mdf}(a)], the relaxation is completed and the system enters into the stationary regime.
Note that in Sec.~\ref{subsec:boltzmann} we have defined the characteristic relaxation time $\tau$ as the time at which $n_k$ develops a maximum at $k=0$ [see the discussion below Eq.~(\ref{nt_poisson})].

\begin{figure}[!t]
\includegraphics[width=.96\columnwidth]{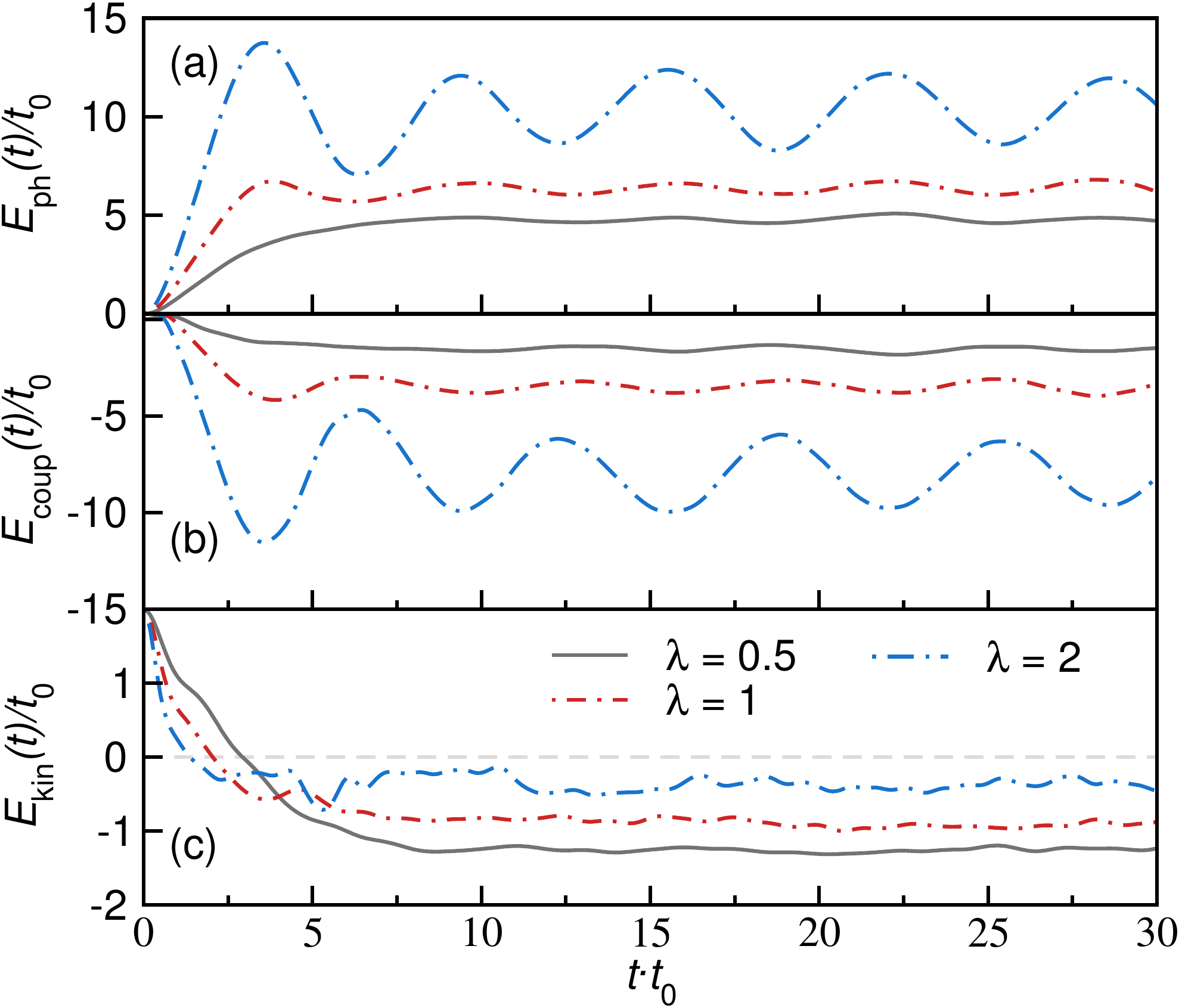}
\caption{(Color online)
Time evolution for the three competing parts of the Hamiltonian: (a) phonon energy $E_{\rm{ph}}(t)$, (b) electron-phonon coupling energy $E_{\rm coup}(t)$, and (c) electronic kinetic energy $E_{\rm kin}(t)$.
We set $\omega_0 = t_0$ and show results using LFS for the three parameter values $\lambda = 0.5$, $1$ and $2$ (we use $L = 12$ for the first two systems and $L = 8$ for the last one).
}
\label{fig:Etime}
\end{figure}

When the electron-phonon coupling is increased, a crossover from the large to the small polaron regime takes place in the ground state of the Holstein model~\cite{fehske2007}.
The dimensionless electron-phonon coupling $\lambda$, Eq.~(\ref{def_lambda}), represents a natural measure of the crossover: $\lambda^* \approx 1$ (the quantitative value depends on the adiabaticity ratio).
Figure~\ref{fig:Etime} shows the time evolution of the three competing terms in the Hamiltonian: $E_{\text{ph}}(t)$, $E_{\text{coup}}(t)$ and $E_{\text{kin}}(t)$ for $\lambda=0.5$, $1$ and $2$ at fixed $\omega_0/t_0 = 1$.
In terms of the energy transfer from the electron to phonons, no drastic features occur at $\lambda \approx 1$, and the relaxation time of the electronic energy $E_{\text{kin}}(t)$ in Fig.~\ref{fig:Etime}(c) continuously decreases with increasing $\lambda$.
Larger differences can be observed in the stationary regime, where oscillations of $E_{\text{ph}}(t)$ and $E_{\text{coup}}(t)$ become noticeable in the strong-coupling regime (this will be discussed in more detail in Sec.~\ref{subsec:steady}).
The momentum distribution function $n_k$ at $\lambda=2$ [see Fig.~\ref{fig:mdf}(b)] also exhibits oscillations in the stationary regime, however, it remains peaked at $k=0$ for most of the time.

Figure \ref{fig:Etime} also addresses a parameter regime that has not been covered by our perturbative analysis in Sec.~\ref{sec:perturbative}:
this is the regime where the electron-phonon coupling is large while the phonon energy is smaller than the electronic bandwidth (c.f. $\lambda=2$ and $\omega_0=t_0$).
In such a case, one can still observe the relaxation regime at short times in the time evolution of the kinetic energy, Fig.~\ref{fig:Etime}(c), which starts at $E_{\rm kin}(t=0) = 2t_0$ and becomes negative at sufficiently large time.
%This is different from the anti-adiabatic regime discussed in Sec.~\ref{sssec:sclimit} where the kinetic energy returns back to its initial value, see Eq.~(\ref{eq:Ekin_antiad}).
In contrast to the kinetic energy, $E_{\text{ph}}(t)$ and $E_{\text{coup}}(t)$ clearly exhibit oscillations with the period $2\pi/\omega_0$ and the entire energy transfer takes place already within a single period.

%%%%%%%%%%%%%%%%%%%%%%%%%%%
\subsubsection{Stationary regime} \label{subsec:steady}

\begin{figure}[!t]
\includegraphics[width=.96\columnwidth]{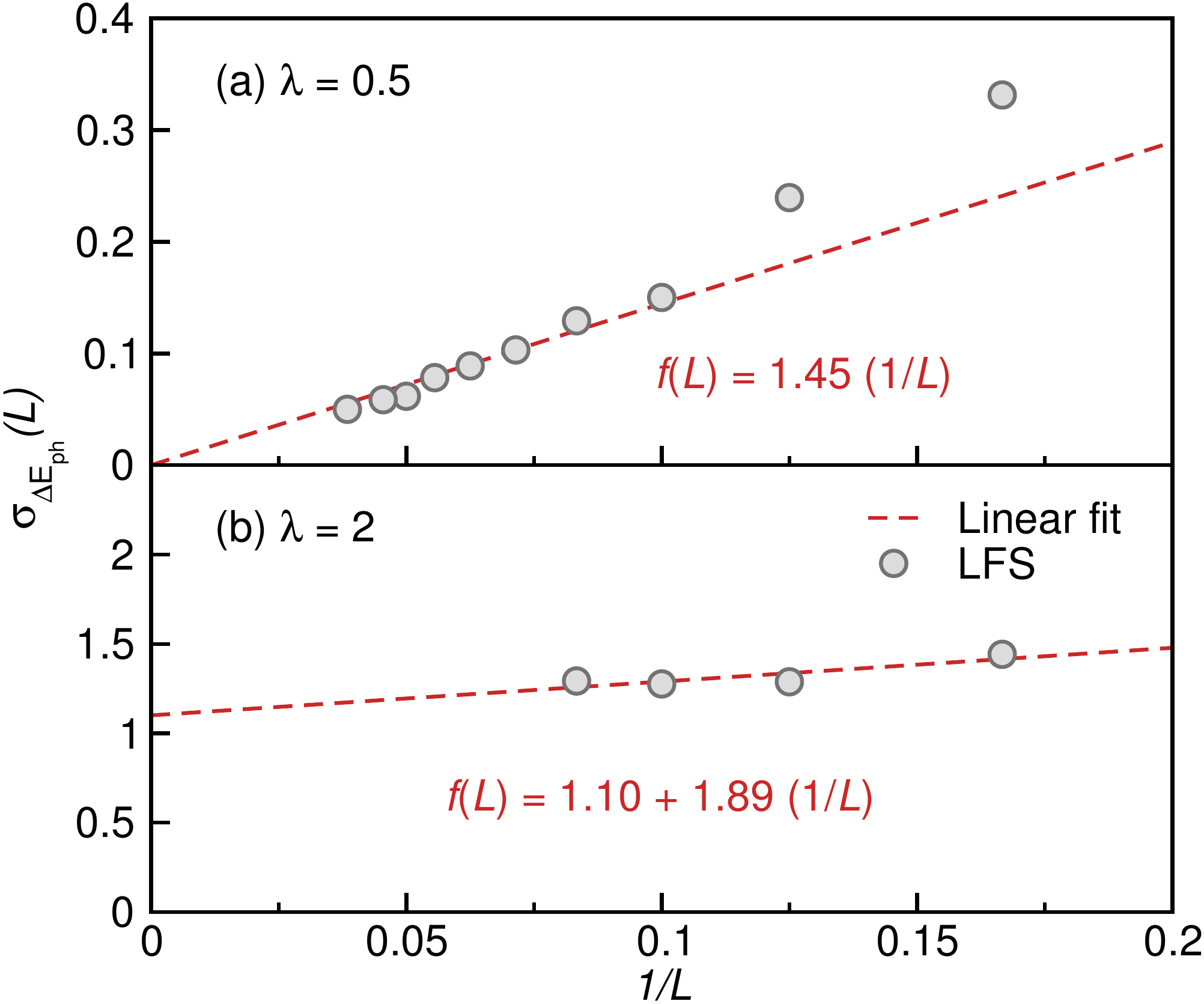}
\caption{(Color online)
Standard deviation $\sigma_{\Delta E_{\rm ph}}$ [see Eq.~\eqref{eq:sigma}] of the phonon energy $E_{\rm ph}(t)$ in the stationary regime, as a function of the inverse lattice size.
Circles represent results using LFS, while the dashed lines in (a) and (b) are the linear fitting functions $f(L)$ with one and two free parameters, respectively.
(a) $\lambda = 0.5$, where oscillations vanish in the thermodynamic limit (the two points for the smallest $L$ were not included in the fit).
(b) $\lambda = 2.0$, where oscillations persist for all $L$.
}
\label{fig:oscdecay}
\end{figure}

In the introductory part, we defined the stationary regime as the regime where no net energy transfer takes place between the electron and the phonon system.
If the time dependence of observables exhibits persistent coherent oscillations, the latter statement corresponds to the average over the oscillation period (in most cases, this equals $2\pi/\omega_0$).

One of the main questions about the oscillations in the stationary regime is whether they persist in the limit $L\to \infty$.
To investigate this, we define the variance of temporal fluctuations of an observable $A(t)$ about the average in the stationary regime
\begin{align}
\sigma_{\Delta A}^2 =\frac{1}{t_2-t_1} \int\limits_{t_1}^{t_2} \left( A(t)-\bar{A} \right)^2 dt, \label{eq:sigma}
\end{align}
where $t_1$ and $t_2$ are chosen to be at the minimum and maximum of the oscillations, and $\bar{A}$ represents the time average.
Figures~\ref{fig:oscdecay}(a) and~\ref{fig:oscdecay}(b) show the standard deviation of temporal fluctuations of the phonon energy at $\lambda = 0.5$ and $\lambda=2$, respectively.
In the first case, the oscillations vanish with the system size, while at $\lambda=2$, the oscillations seem to persist for any lattice size.
Since analytical arguments in Sec.~\ref{sssec:sclimit} yield the time dependence with undamped oscillations in the strong-coupling anti-adiabatic limit $t_0 \to 0$, it is not surprising that away from this limit, the oscillations still govern the dynamics for all times.

\begin{figure}[!t]
\includegraphics[width=.96\columnwidth]{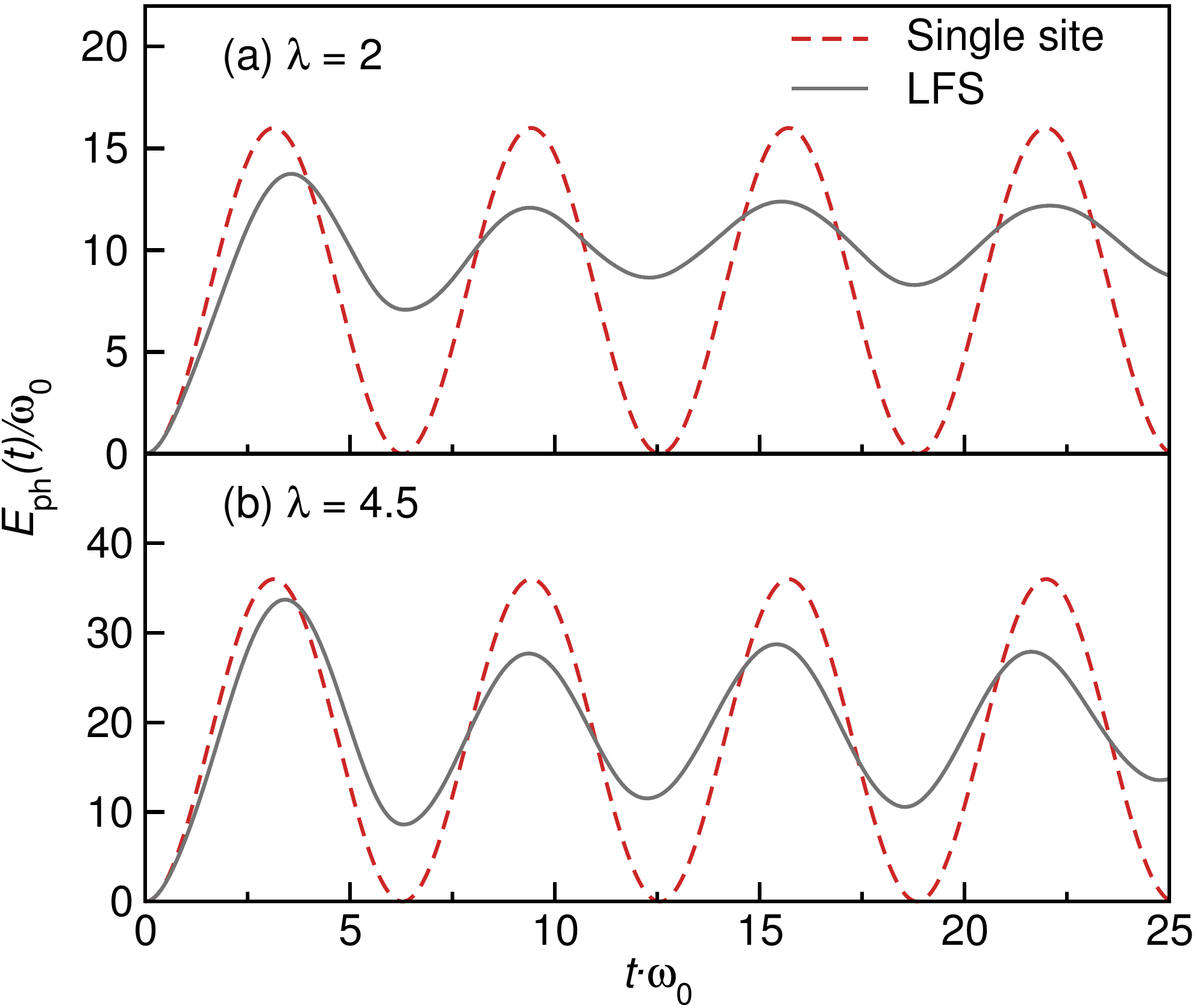}
\caption{(Color online)
Time evolution of the phonon energy $E_{\rm ph}(t)$ in the strong-coupling regime.
The solid lines represent numerical results using LFS for $t_0 = \omega_0$ and $L=8$, while the dashed lines represent the single-site dynamics ($t_0 = 0$), given by Eq.~(\ref{nph_t0limit}).
(a) $\lambda=2$, which corresponds to $g=2$.
(b) $\lambda=4.5$, which corresponds to $g=3$.
}
\label{fig:oscnodecay}
\end{figure}

It is nevertheless still an intriguing question how much information about the dynamics is already included in the solution of the strong-coupling anti-adiabatic limit in Sec.~\ref{sssec:sclimit}.
In Fig.~\ref{fig:Etime}, we have shown that the kinetic energy at $\lambda=2$ is already close to zero in the stationary state.
Because of energy conservation, this implies that the phonon and the coupling energies oscillate with the same phase and a similar amplitude, $E_{\rm ph}(t) = - E_{\rm coup}(t) + E_{\rm total}$, consistent with the result from Eq.~(\ref{nph_t0limit}).
In Figs.~\ref{fig:oscnodecay}(a) and~\ref{fig:oscnodecay}(b) we compare
numerical results for $E_{\rm ph}(t)$ at $t_0 = \omega_0$ to the analytical
result  at $t_0 = 0$ [see Eq.~(\ref{nph_t0limit})] at the same ratio $\gamma/\omega_0$.
We take $\lambda=2$ and $4.5$ in panels (a) and (b), which correspond to $g=2$ and $3$, respectively.
The oscillation period $2\pi/\omega_0$ matches almost perfectly in all cases, suggesting its universality for phonon-related quantities.
However, the amplitude of oscillations at $t_0=0$ (given by $2g^2$) is larger than at $t_0 = \omega_0$.
There are, in fact, two effects when $\lambda$ increases: both the amplitude of oscillations and the average value approach 
the result~(\ref{nph_t0limit}).
In Sec.~\ref{sec:optmodes} we will discuss the optimal modes emerging from the
single-site reduced density matrix, which represent a complementary approach 
to study the structure of phonon modes in non-equilibrium.
We will further confirm that the oscillations at finite $\lambda$ are indeed related to the single-site coherent oscillations characterizing the $t_0\to 0$ limit.

\begin{figure}[!t]
\includegraphics[width=.96\columnwidth]{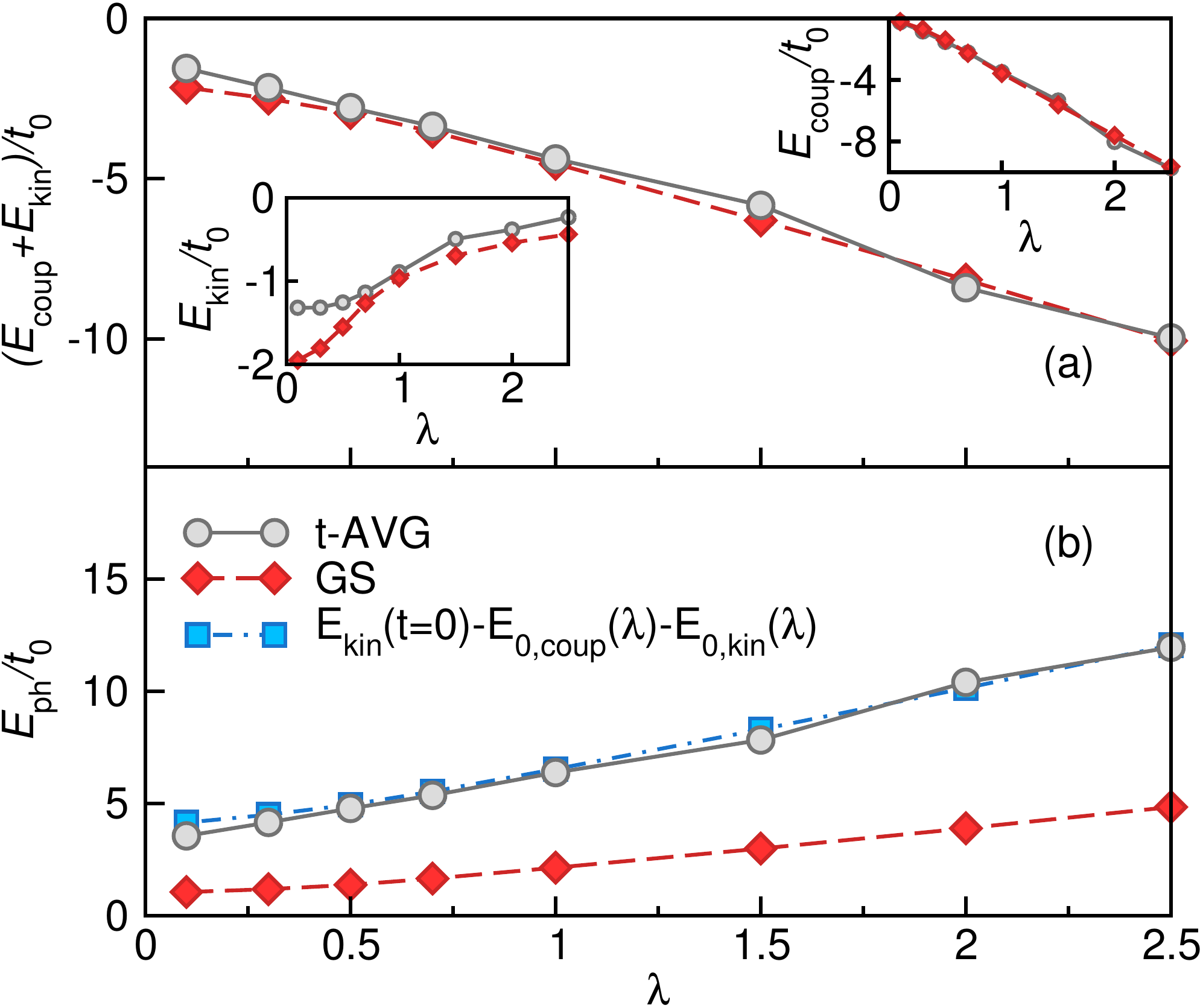}
\caption{(Color online) 
Time-averaged values (t-AVG) of the three competing parts of the Hamiltonian in the stationary state as a function of electron-phonon coupling $\lambda$ at $L=12$ and $\omega_0 = t_0$, obtained by using LFS.
The values are compared to the polaron ground state (GS) energies $E_{\rm kin}$, $E_{\rm coup}$ and $E_{\rm ph}$ at $K=\pi$.
(a) Main panel: $E_{\rm kin} + E_{\rm coup}$ versus $\lambda$ in the ground state (diamonds) compared to the corresponding time-averaged values in the stationary state (circles).
The insets display the comparison of the individual terms.
The data at $\lambda \to 0$ exhibit an $L$-dependence due to the large spatial extent of the polaron.
(b) Phonon energy $E_{\rm{ph}}$ versus $\lambda$.
Since the averages in the stationary state clearly exceed the ground-state values, we also compare to $E_{\rm kin}(t=0)-(E_{0,\rm coup}+E_{0,\rm kin})$,
where $E_{\rm kin}(t=0)= E_{\rm total}$, and $E_{0,\rm coup}$, $E_{0,\rm kin}$ are measured in the ground state.
}
\label{fig:Eavg}
\end{figure}

Finally, we analyze the values of the energy terms $E_{\rm ph}(t)$, $E_{\rm coup}(t)$ and $E_{\rm kin}(t)$ in the stationary regime as a function of electron-phonon coupling $\lambda$ (if the observable oscillates in the stationary regime, we take the average over a sufficiently large time interval).
We also compare these values to their respective ground-state values for the Holstein polaron with the same total momentum, $K=\pi$.
Results are shown in Fig.~\ref{fig:Eavg}.
While kinetic and coupling energies in general relax to values close to their respective ground-state values, this is not the case for the phonon energy [see Fig.~\ref{fig:Eavg}(b)].
Obviously, since the total energy of our closed system exceeds the ground-state energy of the Holstein polaron, it can not happen that all three contributions to the total energy  relax to their ground-state values.
The important message coming from Fig.~\ref{fig:Eavg} is therefore that practically all the excess energy is stored in the form of excess phonons in the stationary state.
The precise quantitative analysis of deviations of the kinetic and the electron-phonon coupling energy with respect to their ground-state values also represents an interesting aspect, which is, however, beyond the scope of this paper.

The latter result allows us to address a more ambitious question:
If we know the energy of the initial state, and if we know the ground-state energies for the system for given model parameters, is it possible to estimate the number of phonons at asymptotically long times without performing the time evolution? Or, in other words, does the numerical data support a simple rule of thumb to 
predict $\bar N_{\rm ph}$?
Figure~\ref{fig:Eavg}(b) shows that such a heuristic estimate is indeed possible.
The number of phonons is, to a good approximation, given by
\begin{equation} \label{heuristic}
\bar{N}_{\rm ph} = \frac{E_{\rm total} - \left  (E_{0,\rm kin} + E_{0,\rm coup} \right )}{\omega_0},
\end{equation}
where $E_{0,\rm kin}$ and $E_{0,\rm coup}$ are the ground-state energy of the kinetic and electron-phonon coupling part of the Hamiltonian.
Even though the latter result appears to be very simple, it contains a potentially non-intuitive aspect:
If the electron-phonon coupling is not small, it is not correct to view the time
evolution solely as the energy transfer from the electron (i.e., the electronic
kinetic energy) to phonons (i.e., the phonon energy $E_{\rm ph}$).
Instead, a considerable part of energy is stored in the coupling energy as well, and the appropriate number of excess phonons can only be obtained if the electronic {\it and} the coupling energy are taken together.
Moreover, Eq.~(\ref{heuristic}) is also helpful for setting up parameters for numerical simulations for which the maximal number of phonons represents the bottleneck for the efficiency.

%%%%%%%%%%%%%%%%%%%%%%%%%%%%
\subsection{Crossover from weak to strong coupling in the anti-adiabatic regime}

\begin{figure}[!t]
\includegraphics[width=.96\columnwidth]{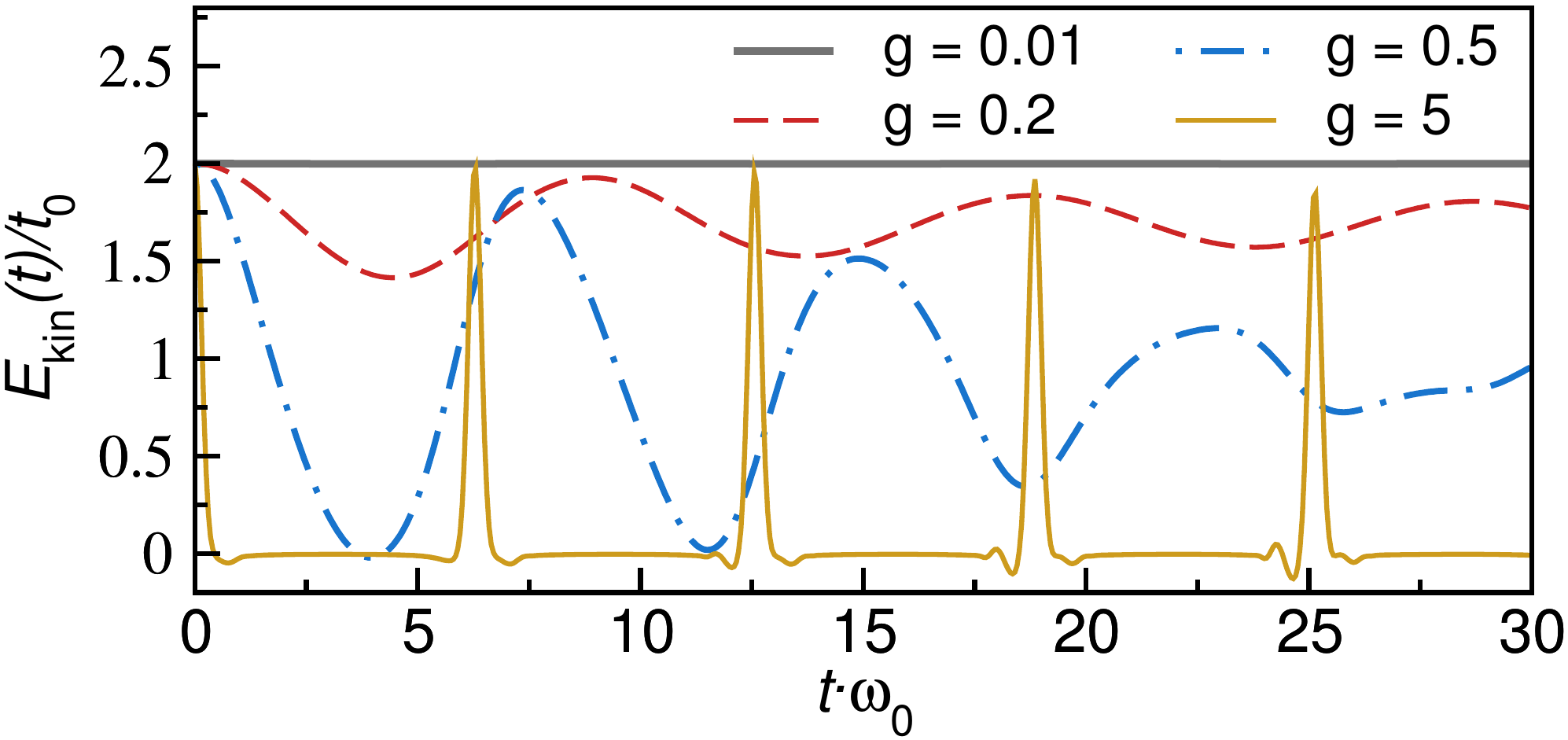}
\caption{(Color online)
The electron kinetic energy $E_{\rm kin}(t)$ at $\omega_0/t_0 = 10$ for several values of the electron-phonon coupling $g = \gamma/\omega_0$.
We use LFS and set $L=12$.
}
\label{fig:Ekin_crossover_omega}
\end{figure}

We complete our investigation by studying the crossover from weak to strong electron-phonon coupling at $\omega_0/t_0=10$.
The evolution of the phonon energy $E_{\rm ph}(t)$ for weak electron-phonon coupling has already been analyzed in Figs.~\ref{fig:weak_antiadia} and~\ref{fig:Eph_lambda}(a).
The response is governed by coherent oscillations without any significant energy redistribution.
This is corroborated by showing $E_{\rm kin}(t)$ in the weak-coupling regime in Fig.~\ref{fig:Ekin_crossover_omega} (see the curves at $g=\gamma/\omega_0=0.01$ and $0.2$), which shows that even the maximal temporal change of $E_{\rm kin}(t)$ is only a tiny fraction of the electronic bandwidth.
In contrast to the case $\omega_0 = t_0$ studied in Sec.~\ref{sec:crosslambda}, therefore, the weak-coupling anti-adiabatic regime is characterized by negligible energy transfer between different parts of the Hamiltonian.

Nevertheless, weak- and strong-coupling anti-adiabatic regimes are different with respect to the temporal evolution since in the latter case the amplitude of oscillations becomes very large.
We show in Fig.~\ref{fig:Ekin_crossover_omega} how $E_{\rm kin}(t)$ changes when $g$ is varied from small ($g \ll 1$) to large values ($g > 1$).
In particular, the strong-coupling regime is governed by oscillations  between the initial value $E_{\rm kin}=2t_0$ and $E_{\rm kin}\approx 0$.
This range of oscillations is consistent with the result from perturbation theory, Eq.~(\ref{eq:Ekin_antiad}), and exhibits the revivals of the initial state at the integer multiples of the phonon period $2\pi/\omega_0$.
These two regimes of oscillations are connected by a crossover regime where the amplitude of oscillations decreases (see the $g=0.5$ curve in Fig.~\ref{fig:Ekin_crossover_omega}) and the reduction of the kinetic energy is of the order of $t_0$.
However, there is no indication of the amplitude of oscillations decay to zero for large $L$ based on the available system sizes.

%%%%%%%%%%%%%%%%%%%%%%%%%%%%%%%%%%%%%%%%%%%%%%%%%%%%%%%%%%%%%%%%%%%%%%%%%%%%%%%
\section{Optimal phonon modes and single-site entanglement entropy} \label{sec:optmodes}

\subsection{Optimal phonon modes}
\label{subsec:optmodes}

As mentioned in the introductory section, the number of phonon states needed to correctly describe an electron-phonon coupled system can be quite large.
For the single-electron problem studied in this paper, we have shown that the limited functional space basis works very efficiently.
However, it is typically restricted to systems that contain only a single (or a few) charge carriers.
Zhang {\it et al.}~first introduced the optimal mode basis for exact diagonalization~\cite{zhang98} with the goal to obtain an efficient truncation scheme for the on-site Hilbert space of bosons that is able to represent the target state with fewer basis states than in the occupation number basis.

A physical example for such an optimal basis can be given for the strong-coupling, anti-adiabatic limit $\gamma, \omega_0~\gg~t_0$.
In this case, the Hamiltonian is well approximated by Eq.~\eqref{sec2:effham}
and only two local phonon states are needed to construct the ground
state of Eq.~(\ref{eq:gs}): the first one is the phonon vacuum state if the site is not occupied by the
electron, and the second one is a coherent state of phonons if the site is occupied by the electron.
Thus two phonon states (for each site) build an optimal basis for the ground state.
Any coherent state $\ket{\beta}$ with $b_j \ket{\beta} = \beta \ket{\beta}$
(for any site $j$) can be expressed in the local phonon occupation number basis 
$\ket{\beta}=\sum_{n=0}^{\infty} \langle n \vert \beta \rangle \ket{n}$.
The weight of each occupation number state is given by a Poisson distribution
\begin{equation} \label{def_poisson}
{\cal P}(n;\vert\beta\vert^2) = \vert \langle n \vert \beta \rangle \vert^2 = \frac{\vert
\beta\vert^{2n} e^{-\vert\beta\vert^2}}{n!} 
\end{equation}
with $n=0,1,2, \dots$.
The average number of local phonons is equal to the variance of this distribution
\begin{align}
N_{{\rm ph}}^{(j)} = \braket{b_j^{\dag} b_j} = \text{Var}[{\cal P}(n;\vert\beta\vert^2)] = \vert \beta\vert^2.
\end{align}
For the ground-state Holstein polaron~(\ref{eq:gs}) we have $\beta=g$ and thus
$N_{\rm ph} = g^2$.
In other words, this ground state can be constructed in a local basis of 
dimension $d=2$ consisting of the phonon vacuum and a coherent phonon state, while one needs a much larger number 
$\propto g^2$ of states in the occupation number basis.
Similarly, we can determine the optimal states for the time-dependent solution 
$\ket{\psi(t)} = e^{-iS} \ket{\widetilde{\psi}(t)}$ 
with $\ket{\widetilde{\psi}(t)}$ given by Eq.~(\ref{eq:solution2}). 
We find that for all times $t$ only two optimal phonon states are required
to describe the state $\ket{\psi(t)}$: the phonon vacuum state and a coherent
state with a complex eigenvalue $\beta = g (1-e^{-i\omega_0t})$.
This corresponds to a number of bare phonon states 
$N_{\rm ph} = 2g^2 [1-\cos(\omega_0t)]$ in agreement with the phonon energy~(\ref{nph_t0limit}).

We next describe how to extract this basis from an arbitrary state.
This is closely related to the procedure in  DMRG
methods~\cite{zhang98,schollwock2005density,schollwock2011density}.
The lattice is split into two parts, say $S$ and $E$, and the reduced density
matrix $\rho_{S}$ of a quantum state $\ket{\psi}$ for the subsystem $S$ is defined by
\begin{align} \label{def_rdm}
\rho_{S} = {\rm tr}_{E} \left( \ket{\psi} \bra{\psi} \right) = \sum\limits_{\alpha} w_{\alpha} \ket{\alpha} \bra{\alpha}.
\end{align}
We get an optimal basis  by selecting the density-matrix eigenstates 
$\ket{\alpha}$ with the highest eigenvalues $w_{\alpha}$.
If the distribution of weights $w_{\alpha}$ is sharp enough, 
the properties of the quantum state $\ket{\psi}$ are almost completely 
determined by the highest optimal states only. 
The optimal phonon basis is a special case where the subsystem $S$ is a single site.

Since the Hamiltonian conserves the number of electrons, the reduced density matrix $\rho_S$ is block-diagonal in the number of electrons $N_{\rm e}$.
In the case of the Holstein model with one electron, there are only two distinct blocks that we denote by $N_{\rm e}=0$ and $N_{\rm e}=1$.
Since we restrict our subsystem $S$ to a single site only, each block has the
dimension $N_{\rm max}+1$ of the phonon occupation number basis (so-called bare basis).
Thus the eigenstates of the reduced density matrix $\rho_S$ can be grouped into
two sets of $N_{\rm max}+1$ states according to the corresponding electron
occupation number.
Thereafter we denote the eigenstates in the
block $N_{\rm e}=0$ by $\ket{\alpha'}$ and those in the block $N_{\rm e}=1$ by $\ket{\alpha''}$.  
In an algorithm that uses this approach for truncating the local basis, 
the number $N_{\rm{opt}}$ of kept states is smaller than the number of bare modes in the basis, 
i.e, $N_{\rm{opt}} < N_{\rm{max}}$.
Here we keep the full optimal basis, 
i.e., $N_{\rm{opt}} = N_{\rm{max}}$.

The sum rule for the weights $\omega_\alpha$ coming from the two blocks can then be reorganized as
\begin{equation}
1 = \sum_{\alpha=0}^{2N_{\rm max}+1} w_\alpha = \sum_{\alpha'=0}^{N_{\rm max}}
w_{\alpha'} + \sum_{\alpha''=0}^{N_{\rm max}}  w_{\alpha''}
\end{equation}
with the partial sum rules
\begin{eqnarray}
\sum_{\alpha'} w_{\alpha'} & = &  \frac{L-1}{L} \\
\sum_{\alpha''} w_{\alpha''} & = & \frac{1}{L} \label{weights_block1}
\end{eqnarray}
representing the probability of finding no electron or one electron on a given site, respectively.

\subsection{Von Neumann entropy}
\label{subsec:svn}

A convenient measure of the entanglement between a single site and the rest of the system is the von Neumann entropy $S_{\rm vN}$.
It is defined as
\begin{align}
S_{\rm vN} = - \sum\limits_{\alpha=0}^{2N_{\rm max}+1} w_{\alpha} {\rm ln}\left( w_{\alpha} \right),
\end{align}
where the $w_\alpha$ are the eigenvalues of the single-site reduced density matrix $\rho_S$, see Eq.~(\ref{def_rdm}).
In our work we use the von Neumann entropy to get insight into the increase of the number of relevant optimal modes: a larger entanglement entropy indicates that more optimal modes are relevant for the dynamics.
In a more general context, since the excess energy is an intensive quantity, this classifies our set-up as a so-called local quench problem~\cite{eisler07,calabrese07}.

For our initial state~(\ref{psi0}) there is no phonon in the system, hence the reduced density matrix contains only two non-zero entries, $w_0 = \frac{L-1}{L}$ and $w_1 = \frac{1}{L}$, which correspond to the zero-phonon state on a site without and with the electron, respectively.
The initial entropy is therefore
\begin{equation} \label{svn0}
S_{\rm vN} = \frac{1}{L} {\rm ln}(L) + \frac{L-1}{L} {\rm ln}\left( \frac{L}{L-1} \right).
\end{equation}
The entropy is hence finite because of the block structure of the reduced density matrix.
At the same time, this entropy also represents the minimal value of $S_{\rm vN}$ in a delocalized electronic system.

%%%%%%%%%%%%%%%%%%%%%%%%%%%%%%%%%
\subsection{Results}
\label{subsec:optmodes_results}

\begin{figure}[!t]
\includegraphics[width=.96\columnwidth]{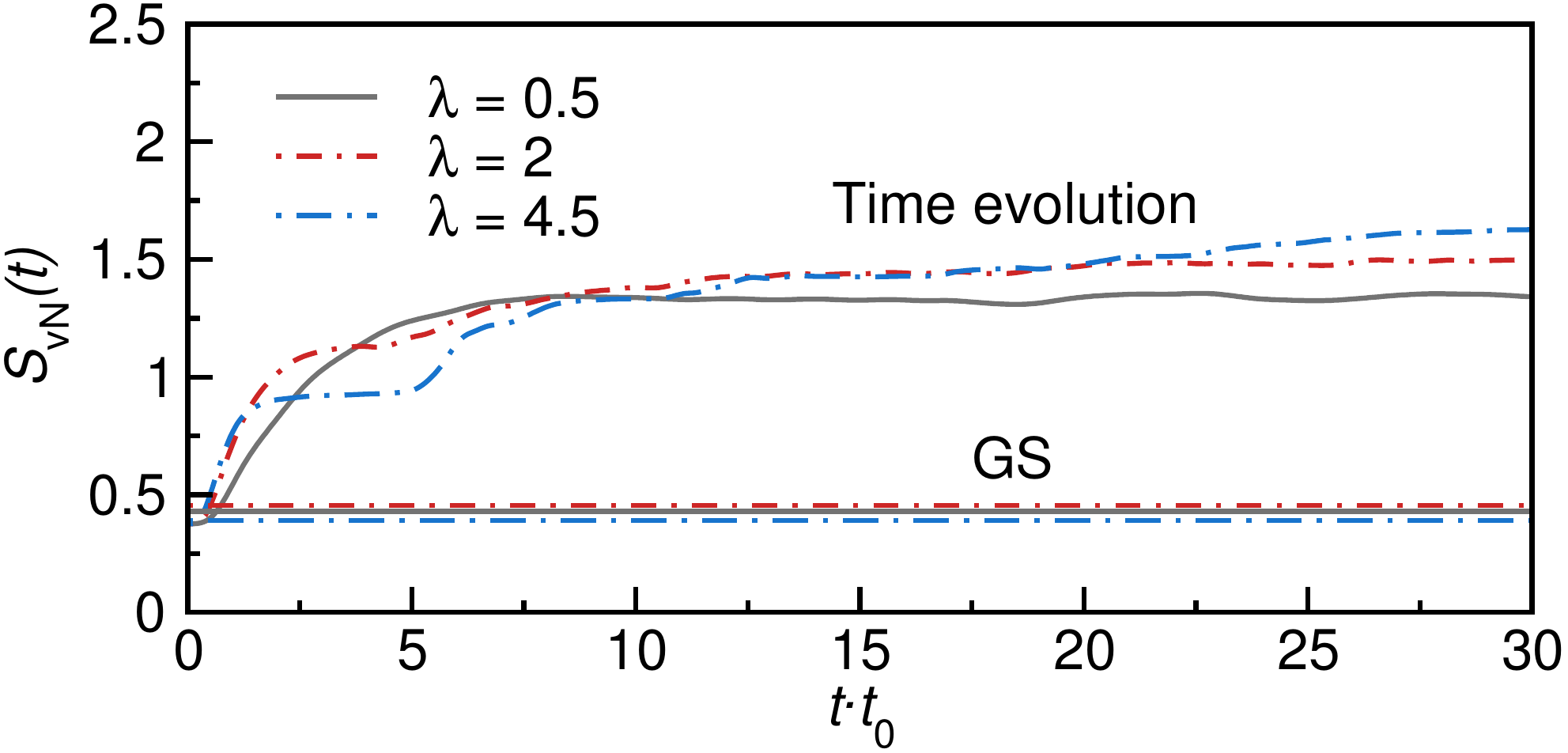}
\caption{(Color online)
Time evolution of the von Neumann entropy $S_{\text{vN}}(t)$ for  $\omega_0 = t_0$ and different values of the electron-phonon coupling $\lambda=0.5$, $2$ and $4.5$.
The values at $t=0$ are determined by Eq.~(\ref{svn0}).
The horizontal lines represent the corresponding ground-state (GS) values. 
We use LFS with $L=8$.
}\label{fig:svn}
\end{figure}

We now turn to the discussion of numerical results for the optimal modes and the von Neumann entropy, obtained by time evolving the initial wavefunction~(\ref{psi0}) within LFS.
We first show the results for the time dependence of the von Neumann entropy $S_{\rm vN}(t)$ in Fig.~\ref{fig:svn} for three different values of the electron-phonon coupling $\lambda=0.5$, $2$ and $4.5$.
It starts from the minimum value at $t=0$, given by Eq.~(\ref{svn0}), and it monotonically increases until it reaches the stationary regime.
The initial slope of $S_{\rm vN}(t)$ increases with $\lambda$, in apparent correlation with the initial slope of $E_{\rm kin}(t)$ shown in Fig.~\ref{fig:Etime}(c).
Interestingly, even though in the stationary regime at $\lambda=2$ the phononic energy $E_{\rm ph}(t)$ and the electron-phonon coupling energy $E_{\rm coup}(t)$ exhibit coherent oscillations [c.f.~Figs.~\ref{fig:Etime}(a) and~\ref{fig:Etime}(b)] with a non-vanishing amplitude in the thermodynamic limit [c.f.~Fig.~\ref{fig:oscdecay}(b)], the von Neumann entropy barely exhibits any temporal dependence.
Since $S_{\rm vN}(t)$ measures the entanglement entropy between a single site and the rest of the system, this result supports the view that the coherent oscillations indeed stem from the single-site dynamics discussed in Sec.~\ref{sssec:sclimit}.
The time evolution at very strong coupling $\lambda=4.5$ does not reach a stationary value within $t t_0 < 30$, which may heuristically be understood within the small-polaron picture, where the effective hopping amplitude of the composite electron and the on-site phonon cloud is much smaller than the bare hopping amplitude $t_0$.

\begin{figure}[!t]
\includegraphics[width=.96\columnwidth]{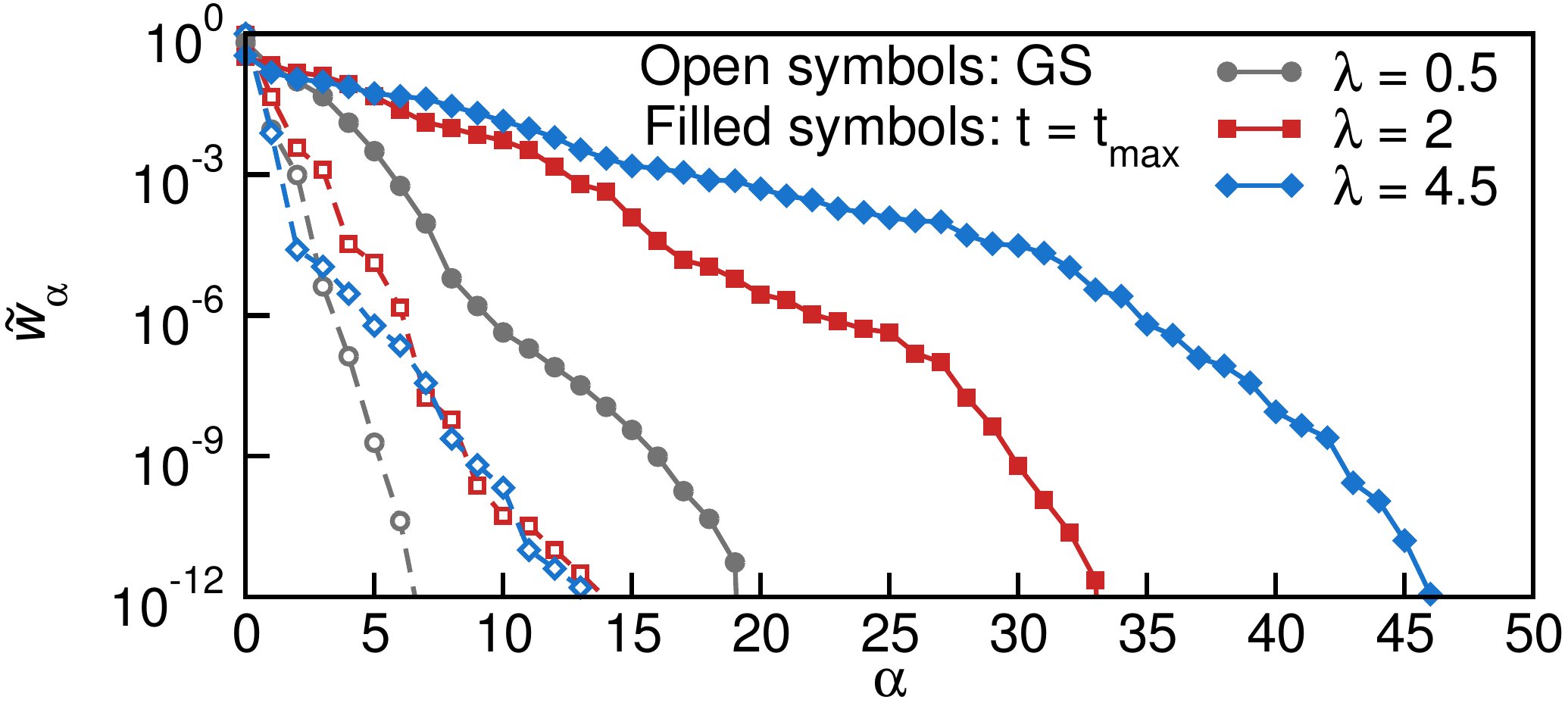}
\caption{(Color online)
Weights $\widetilde w_\alpha$ of the optimal modes in descending order for  $\omega_0 = t_0$ and different values of the electron-phonon coupling $\lambda=0.5$, $2$ and $4.5$ (we use LFS with $L=12$ for $\lambda=0.5$ and $L=8$ for the latter two systems).
Filled symbols represent weights in the stationary regime while open symbols represent the corresponding ground-state (GS) values.
We choose a time $t_{\rm max}$ in the stationary regime at which the phonon energy $E_{\rm ph}(t)$ has a local maximum (c.f.~Figs.~\ref{fig:Etime}(a) and~\ref{fig:oscnodecay}), i.e.,
$t_{\rm max} t_0 = 15.8$ for $\lambda=0.5$ and $t_{\rm max} t_0 = 9.3$ for $\lambda=2$ and $4.5$. 
}
\label{fig:weight}
\end{figure}

\begin{figure}[!t]
\includegraphics[width=.96\columnwidth]{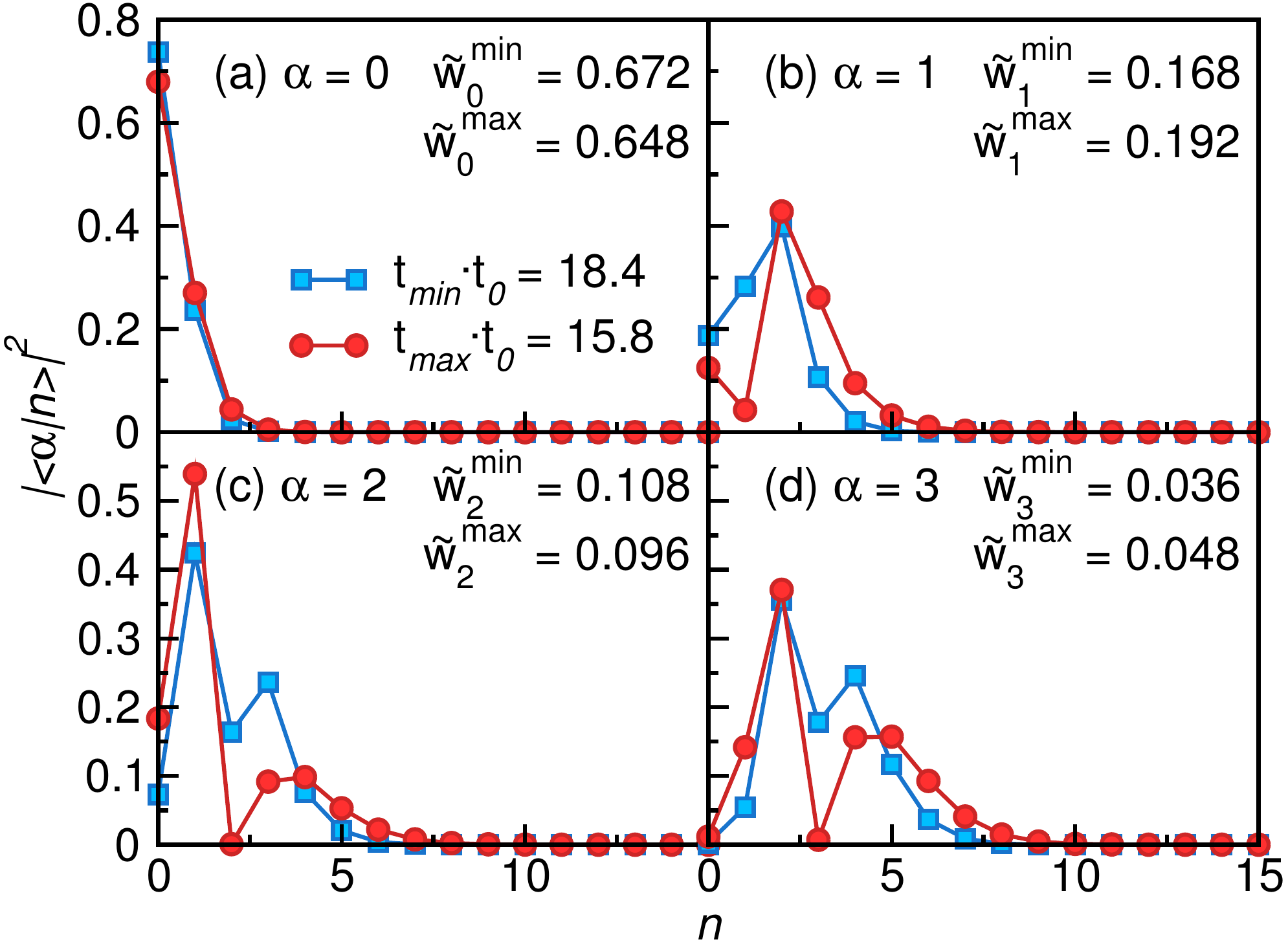}
\caption{(Color online)
Components $\vert \langle n \vert \alpha \rangle \vert ^2$ 
of the four highest optimal modes $\ket{\alpha}$ in the phonon occupation number basis $\{ \ket{n} \}$.
Results are shown in the stationary regime using LFS at $\lambda=0.5$, $\omega_0 = t_0$ and $L=12$ for
 two different times: at $t_{\rm min}$ (squares) the phonon energy $E_{\rm{ph}}(t)$ has a local minimum, while at $t_{\rm max}$ (circles) it has a local maximum [see the time evolution of  $E_{\rm{ph}}(t)$ in Fig.~\ref{fig:Etime}(a)].
$\widetilde{w}_\alpha^{\rm min}$ and $\widetilde{w}_\alpha^{\rm max}$ represent the optimal mode weights~(\ref{def_tildew})
at the given times. 
}
\label{fig:omodeweak}
\end{figure}

\begin{figure}[!t]
\includegraphics[width=.96\columnwidth]{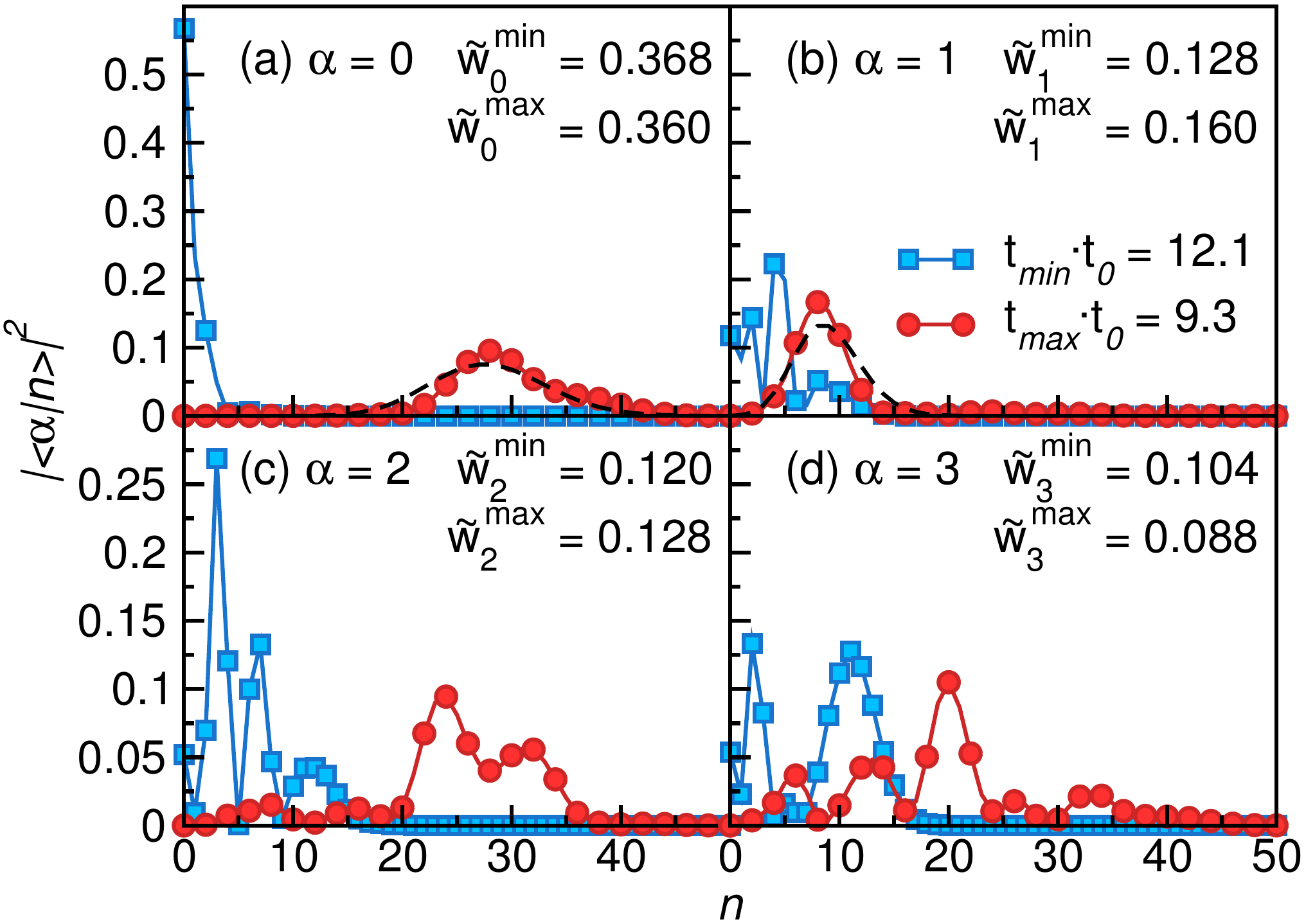}
\caption{(Color online)
Components $\vert \langle n \vert \alpha \rangle \vert ^2$ of the four highest
optimal modes $\ket{\alpha}$ in the phonon occupation number basis $\{ \ket{n} \}$.
Results are shown in the stationary regime using LFS at $\lambda=4.5$, $\omega_0 = t_0$ and $L=8$
for two different times: at $t_{\rm min}$ (squares) the phonon energy $E_{\rm{ph}}(t)$ has a local minimum, while at $t_{\rm max}$ (circles) it has a local maximum [see the time evolution of  $E_{\rm{ph}}(t)$ in Fig.~\ref{fig:oscnodecay}(b)].
$\widetilde{w}_\alpha^{\rm min}$ and $\widetilde{w}_\alpha^{\rm max}$ represent the optimal mode weights~(\ref{def_tildew})
at the given times.
Dashed lines in (a) and (b) represent Poisson distributions~(\ref{def_poisson}).
}
\label{fig:omodestrong}
\end{figure}

For the remainder of the study, we focus on the stationary regime only and analyze the eigensystem of the single-site reduced density matrix $\rho_S$ defined in Eq.~(\ref{def_rdm}).
In particular, we study the eigensystem on the site occupied by the electron, i.e., the $N_{\rm e}=1$ block of $\rho_S$.
Since the corresponding eigenvalues $w_{\alpha''}$ do not sum up to one [see Eq.~(\ref{weights_block1})], we normalize them by a factor  $L$, which is the inverse of the probability to find the electron on a given site.
To simplify the presentation of our results in Figs.~\ref{fig:weight}-\ref{fig:omodestrong} we use the notation
\begin{equation} \label{def_tildew}
\widetilde{w}_{\alpha} = L \cdot w_{\alpha''}, \; \; \ket{\alpha} = \ket{\alpha''}_{N_{\rm e}=1}.
\end{equation}

Figure~\ref{fig:weight} shows the weights $\widetilde{w}_\alpha$ in decreasing order for  three different values of the coupling $\lambda=0.5$, $2$ and $4.5$.
For each $\lambda$, we choose a time $t_{\rm max}$ in the stationary regime where the oscillating phonon energy $E_{\rm ph}(t)$ has a local maximum [c.f.~Figs.~\ref{fig:Etime}(a) and~\ref{fig:oscnodecay}].
We also compare the weights to the corresponding ground-state values.
In general, both in the ground state and in the stationary state, the weights
$\widetilde{w}_\alpha$ decay roughly exponentially (in fact, in Fig.~\ref{fig:weight} that shows that data at $t_{\rm max}$, there is a crossover from a slower to a faster decay).
This suggests that the concept of optimal modes could be used to reduce the
computational cost in correlated electron-phonon systems in non-equilibrium as
already done for ground states and dynamical correlation functions~\cite{zhang98,zhang99}.
Nevertheless, the decay of the weights $\widetilde{w}_\alpha$ is much more rapid in the ground state, consistent with a much smaller entanglement entropy.
In addition, in the stationary regime, the weights decay much slower for larger electron-phonon coupling $\lambda$.
Since the weights are normalized, the smaller decay constant for larger
$\lambda$ indicates that more eigenstates (optimal modes) need to be taken into
account to get the same truncation error, which in turn would make the numerical
computation more demanding.

Finally, we focus on the structure of the optimal modes $\ket{\alpha}$ in the stationary regime.
We express their components $\vert \langle n \vert \alpha \rangle \vert ^2$ 
in the local phonon occupation number basis $\{ \ket{n} \}$ and plot them at two different times $t_{\rm max}$ and $t_{\rm min}$ corresponding to a maximum and a minimum of the phonon energy $E_{\rm{ph}}(t)$, respectively.
Results for the four highest weighted optimal modes are shown in Fig.~\ref{fig:omodeweak} for $\lambda=0.5$ and in 
Fig.~\ref{fig:omodestrong} for $\lambda=4.5$.

In the weak-coupling case, Fig.~\ref{fig:omodeweak}, the modes at  $t=t_{\rm max}$ and $t=t_{\rm min}$  look very much alike, which can be related to the fact that the oscillations are very weak in this regime and even fully vanish in the thermodynamic limit (c.f.~Fig.~\ref{fig:oscdecay}).
In addition, the modes are clearly peaked and formed only by a few bare modes (i.e., occupation number states).
This is expected in the weak-coupling regime where only a relatively small amount of phonons is excited.
The highest weighted mode $\ket{\alpha =0}$, shown in Fig.~\ref{fig:omodeweak}(a), has its peak at zero phonons, but
the second and third highest optimal modes $\ket{\alpha=1}$ and $\ket{\alpha=2}$ shown in Figs.~\ref{fig:omodeweak}(b) and~\ref{fig:omodeweak}(c) seem to have swapped rank because
 their components peak  at the positions of the two- and one-phonon states, respectively.

On the other hand, in the strong-coupling case, the optimal modes at $t=t_{\rm max}$ and $t= t_{\rm min}$ differ significantly.
Figure~\ref{fig:omodestrong} shows results for $\lambda=4.5$ (i.e., $g=3$).
The most striking difference can be observed for the highest weighted mode $\ket{\alpha=0}$ in Fig.~\ref{fig:omodestrong}(a).
At $t=t_{\rm min}$  it is essentially equal to the bare mode with zero phonon
occupation (i.e., the phonon vacuum), while 
at $t=t_{\rm max}$ it strongly resembles a Poisson distribution ${\cal P}(n;\vert \beta \vert^2)$
with a width set by the actual number of phonons in the system, i.e., $\vert
\beta \vert^2 = N_{\rm ph} = E_{\rm ph}(t_{\rm max})/\omega_0$
[compare the dashed line with circles in Fig.~\ref{fig:omodestrong}(a)].
This actual value of $N_{\rm ph}$ differs significantly from the value $\vert \beta \vert^2=4g^2$ 
in the limit $t_0=0$ as seen in Fig.~\ref{fig:oscnodecay}(b).
Nevertheless, this result agrees qualitatively with the picture of a coherent phonon state 
obtained in the $t_0=0$ limit in Sec.~\ref{subsec:optmodes}.
Therefore, a similar dynamics can be observed for the optimal mode with the
largest weight in a system with finite parameters $\lambda=4.5$ and $\omega_0 = t_0$
and in the anti-adiabatic strong-coupling limit.

Apart from the highest optimal mode, also the second highest optimal mode $\ket{\alpha=1}$ shown in Fig.~\ref{fig:omodestrong}(b) exhibits intriguing properties.
At $t_{\rm max}$, it represents a distribution peaked at some finite phonon number.
We plot  the Poisson distribution ${\cal P}(n;g^2=9)$ as a dashed line in Fig.~\ref{fig:omodestrong}(b), 
which corresponds to the ground-state phonon distribution at $t_0=0$.
Interestingly, the latter function strongly resembles the numerically calculated optimal mode.
It suggests that at the local maximum of $E_{\rm{ph}}(t)$, there is a coexistence of distinct coherent states. 
On the other hand, the modes with lower weights shown in Figs.~\ref{fig:omodestrong}(c) and~\ref{fig:omodestrong}(d) 
apparently do not exhibit any simple structure.

%%%%%%%%%%%%%%%%%%%%%%%%%%%%%%%%%%%%%%%%%%%%%%%%%%%%%%%%%%%%%%%%%%%%%%%%%%%%%%%
\section{Summary, Discussion, Conclusion} \label{sec:conclusion}

In summary, we performed a comprehensive analysis of the non-equilibrium dynamics in the Holstein polaron model.
We considered the initial state where the electron is highly excited while the phonon energy is zero.
Two qualitatively different limiting scenarios exist depending on whether the phonon energy is much smaller or much larger than the electronic bandwidth:
in the first one, relaxation sets in, i.e., there is a net energy transfer from the electron to the phonon system,
while in the second case, the response of the system is usually governed by persistent coherent oscillations.
The analysis of the real-time dynamics in the entire parameter regime was done using advanced numerical methods:
by comparing with exact diagonalization and analytical results in limiting cases, 
we benchmarked the TEBD method and diagonalization in LFS.
The increase of the phonon number as a function of time, which typically happens in the transient
dynamics, is the main problem preventing the simulation of arbitrarily long times
for all parameter regimes. Nonetheless, if this initial increase can be correctly captured, then  
the LFS method is  very efficient in describing the
dynamics for very long times on relatively large lattices and in different parameter regimes.
 
The dynamics at weak electron-phonon coupling and small phonon energy compared to the electronic bandwidth is very instructive:
the electron dissipates its energy in the relaxation regime, and then enters the stationary regime where temporal fluctuations about the average value vanish in the thermodynamic limit.
The relaxation is characterized by the redistribution of electronic momenta from
the top ($k=\pi$) towards the bottom ($k = 0$) of the electronic band.
This represents a prototypical example of dissipative dynamics in a closed quantum system.
As one of the main advantages, it allows for a direct comparison of the dynamics emerging from unitary time evolution (calculated using numerical algorithms) with semi-classical approaches.
We calculated the relaxation dynamics from the Boltzmann equation and obtained good agreement with the numerical data.
The results show that the relaxation of the many-body system can well be described by a linear decrease of the electronic
kinetic energy with time, and the characteristic relaxation time is determined by the interplay of all three parameters of the Holstein model.
Assuming a constant electronic density of states, a compact expression for the relaxation time $\tau$ follows from the Boltzmann equation,
$\tau \omega_0 = (16/\pi) (t_0^2/\gamma^2)$.
Our numerical results for relaxation, obtained by using the actual density of
states of a 1D tight-binding system, are interestingly consistent with this prediction.
As a simple consequence, we observe that there is a lower bound for the relaxation time in this parameter regime: it is set by the inverse phonon energy, i.e., $\tau \omega_0 >1$.
Here, the relaxation time $\tau$ represents the complete quasi-particle relaxation time.
It differs qualitatively from the characteristic time $\widetilde \tau$ of a single transition, i.e., a process where a single phonon is emitted: in the latter case, the characteristic transition time for the constant density of states is given by
$\widetilde \tau \omega_0 = (4/\pi) (t_0 \omega_0/\gamma^2)$. 

When the electron-phonon coupling or the phonon frequency are increased, coherent temporal oscillations are enhanced and it becomes more difficult to disentangle the relaxation regime from the stationary regime.
We discussed two well-defined regimes of model parameters where persistent coherent oscillations govern the dynamics:
 when $\omega_0$ is much larger than the electronic bandwidth
and when the electron-phonon coupling $\gamma$ gets much larger than the hopping amplitude $t_0$.
The simplest model which captures the essence of both scenarios is the single-site Holstein model (i.e., when $t_0=0$).
Its ground state is a shifted harmonic oscillator and analytical expressions for the time evolution of observables can be obtained for all times.
This limiting case unveils two important scales for dynamics:
the period of oscillations, which equals $2\pi/\omega_0$, and the amplitude of oscillations, given by $\gamma/\omega_0$.
For our initial state without any phonon, the interpretation of the single-site dynamics is particularly simple:
the phonon state of an occupied site is a coherent state with the
time-dependent eigenvalue $\beta = g(1-e^{-i\omega_0t})$.
As a consequence, one can set an upper bound for the number of emitted phonons in the time evolution: it can not exceed
four times the value found in the ground state.
Our numerical results suggest that this criterion represents a reasonable estimate for the dynamics at finite $t_0$ as well.

We also calculated the time dependence of the single-site reduced density matrix:~it allows for the investigation of the optimal phonon modes and the entanglement entropy during the time evolution.
We showed that the structure of the optimal modes carries valuable information about the dynamics of the system, and hence complements the investigation of non-equilibrium dynamics based on observables only.
In particular, for strong electron-phonon coupling, the structure of the highest-weighted optimal mode resembles the phonon distribution in the $t_0=0$ limit.
This is consistent with the numerical observation that the single-site coherent oscillations, which can be described
analytically for $t_0=0$, persist for finite electron-phonon coupling and finite hopping $t_0$.
On the level of observables, the persistent oscillations can be seen, e.g., when measuring the phonon energy: our numerical results suggest that the amplitude of these oscillations does not vanish in the thermodynamic limit.
The structure of other optimal modes also exhibit interesting, yet less well understood properties.
In any case, non-equilibrium optimal modes represent a promising concept
for characterizing the dynamics as well as for designing novel numerical algorithms~\cite{zhang98} to treat strongly correlated electron-phonon systems out of equilibrium, and hence deserve more attention in future studies.

\section*{Acknowledgement} \label{sec:ack}

We acknowledge stimulating discussions with Janez Bon\v ca, Martin Eckstein, Marcin Mierzejewski and Ines de Vega.
We thank Volker Meden for comments on the previous version of a manuscript.
C.B., F.D., F.H.-M., and E.J. acknowledge support from the DFG (Deutsche Forschungsgemeinschaft) through 
grants Nos.~JE~261/2-1 and HE~5242/3-1 in the Research Unit 
\textit{Advanced Computational Methods for Strongly Correlated Quantum Systems} (FOR 1807).
L.V. is supported by the Alexander von Humboldt Foundation.

%%%%%%%%%%%%%%%%%%%%%%%%%%%%%%%%%%%%%%%%%%%%%%%%%%%%%%%%%%%%%%%
\appendix

%%%%%%%%%%%%%%%%%%%%%%%%%%%%%%%%%
\section{Matrix elements for time evolution in the weak-coupling regime} \label{sec:app1}

In Sec.~\ref{subsec:weakeph}, we obtained analytical expressions for the time dependence of observables in the limit of weak electron-phonon interaction.
Contributions up to ${\cal O}(\gamma^2)$ consist of the three terms given by Eq.~(\ref{eq_u0u1}),~(\ref{eq_u0u2}) and~(\ref{eq_u1u1}).
Here we provide the corresponding matrix elements for the individual parts of the Holstein Hamiltonian, Eq.~(\ref{holstein_ham}), as well as for the momentum distribution function $n_k$.

For the kinetic energy $E_{\rm kin}(t)$, only the second-order terms are non-zero,
\begin{eqnarray}
R_{K,q}^{(1)} & = & 0 \\
R_K^{(2a)} & = & \epsilon_K = -2t_0 \cos{K} \\
R_{K,q}^{(2b)} & = & \epsilon_{K-q} = -2t_0 \cos{(K-q)},
\end{eqnarray}
and the same holds for the momentum distribution function $n_k$
\begin{eqnarray}
R_{K,q}^{(1)} & = & 0 \\
R_K^{(2a)} & = & \delta_{k,K} \\
R_{K,q}^{(2b)} & = & \delta_{k,K-q}.
\end{eqnarray}
For the phonon energy $E_{\rm ph}(t)$, only a single second-order term is non-zero
\begin{eqnarray}
R_{K,q}^{(1)} & = & 0 \\
R_K^{(2a)} & = & 0 \\
R_{K,q}^{(2b)} & = & \omega_0,
\end{eqnarray}
while for the electron-phonon coupling energy $E_{\rm coup}(t)$, which is the only off-diagonal operator in the momentum-space basis, there is a non-zero term already in the first order,
\begin{eqnarray}
R_{K,q}^{(1)} & = & - \frac{\gamma}{\sqrt{L}} \\
R_K^{(2a)} & = & 0 \\
R_{K,q}^{(2b)} & = & 0.
\end{eqnarray}
We apply these values to study dynamics in Sec.~\ref{subsec:intpicture} and~\ref{subsec:boltzmann}.

%%%%%%%%%%%%%%%%%%%%%%%%%%%%%%%%%%%%%%%%   Bibliography

%\bibliographystyle{apsrev}
\bibliography{references}

\end{document}